\documentclass[prd,preprint,tightenlines,floatfix,showpacs,preprintnumbers,nofootinbib,eqsecnum,superscriptaddress]{revtex4}

 \usepackage[dvips,final]{graphicx}
  \usepackage{amssymb}
   \usepackage{amsmath}
    \usepackage{amsfonts}
     \usepackage{epsfig}
      \usepackage{bm}

\usepackage[section]{placeins}

\usepackage{multirow}
\usepackage{ctable}
\usepackage{booktabs}
\usepackage{array}
\usepackage{tabularx}
\usepackage{xcolor}
\usepackage{pstricks}

\newcommand{\bp}{\mbox{\boldmath $p$}}
\newcommand{\bq}{\mbox{\boldmath $q$}}

\newcommand{\aem}{{\alpha_{\mathrm{em}}}}
\newcommand{\ket}[1]{| {#1} \rangle}
\newcommand{\bra}[1]{\langle {#1} |}

\def\lsim{\mathrel{\rlap{\lower4pt\hbox{\hskip1pt$\sim$}}
    \raise1pt\hbox{$<$}}}         
\def\gsim{\mathrel{\rlap{\lower4pt\hbox{\hskip1pt$\sim$}}
    \raise1pt\hbox{$>$}}}         

\begin{document}

\vfill
\title{Two-photon dilepton production in proton-proton collisions: \\
 two alternative approaches}

\author{Marta {\L}uszczak}
\email{luszczak@univ.rzeszow.pl} \affiliation{University of Rzesz\'ow, PL-35-959 Rzesz\'ow, Poland}

\author{Wolfgang Sch\"afer}
\email{Wolfgang.Schafer@ifj.edu.pl} \affiliation{Institute of Nuclear Physics PAN, PL-31-342 Cracow, Poland}

\author{Antoni Szczurek}
\email{antoni.szczurek@ifj.edu.pl} \affiliation{Institute of Nuclear Physics PAN, PL-31-342 Cracow, Poland and\\
University of Rzesz\'ow, PL-35-959 Rzesz\'ow, Poland}

\date{\today}

\begin{abstract}
We investigate different methods to incorporate the effect of 
photons in hard processes. 
We compare two different approaches used for calculating
cross sections for two-photon $p p \to l^+ l^- X$ process. 
In one of the approaches photon is treated as a collinear parton in 
the proton. In the second approach recently proposed a $k_T$-factorization
method is used. We discuss how results of the collinear parton model
depend on the initial condition for the QCD evolution
and discuss an approximate treatment where photon is excluded
from the combined QCD-QED evolution.
We demonstrate that it is not necessary to put photon into the
evolution equation as often done recently but it is sufficient 
to use a simplified approach in which photon couples to quarks 
and antiquarks which by themselves undergo DGLAP evolution equations.
We discuss sensitivity of the results to the choice
of structure function parametrization and experimental cuts in the $k_T$-factorization
approach.
A new optimal structure function parametrization is proposed.
We compare results of our calculations with recent experimental data
for dilepton production and find that in most cases
the contribution of the photon-photon mechanism is rather small.
We discuss how to enhance the photon-photon contribution.
We also compare our results to those of recent measurements of 
exclusive and semi-exclusive $e^+ e^-$ pair production with certain 
experimental data by the CMS collaboration.
\end{abstract}



\maketitle

\section{Introduction}

The two-photon processes may lead to production of two charged
leptons and therefore compete with other sources of dileptons, such
as continuum Drell-Yan processes or resonant production
of vector quarkonia or $Z^0$ boson, which produce dilepton pairs
of large invariant masses. 
Earlier studies of lepton pair production via $\gamma\gamma$ fusion
in inelastic proton-proton collisions can be found in 
\cite{Chen:1973mv,Carimalo:1978bu,Schrempp:1980zx,daSilveira:2014jla}.
For a general review of the $\gamma \gamma$-fusion mechanism, 
see \cite{Budnev:1974de}. Inelastic processes are also included in the 
Monte-Carlo generator LPAIR based on \cite{Vermaseren:1982cz}.

At high energies and small dilepton transverse momenta also 
semi-leptonic decays of pair-wise produced charmed $D$ mesons may 
be an important ingredient of dileptons \cite{Maciula:2010yw}.
Actual contribution of different processes depends strongly on the details
of experimental cuts.

The color singlet exchange of photons naturally leads to rapidity gaps.
If the rapidity veto on particles close to the $l^+ l^-$ vertex
is imposed in addition one can enhance the relative contribution
of the $\gamma \gamma$ processes compared to the QCD Drell-Yan 
mechanism \cite{CMS_gammagamma}. 
The invariant mass distribution of dileptons 
produced in the Drell-Yan processes can be calculated in
collinear-factorization approach (see e.g. the textbook \cite{Ellis:1991qj}).
If one wants to address more differential distributions, say in 
transverse momentum of the lepton pair, one can turn to
$b$-space resummation \cite{Collins:1984kg} or, especially in
the small-$x$ kinematics, $k_T$-factorization 
(see \cite{Szczurek:2008ga,NNS2013,Baranov:2014ewa} for instance).

\begin{figure}
\begin{center}
\includegraphics[width=5cm]{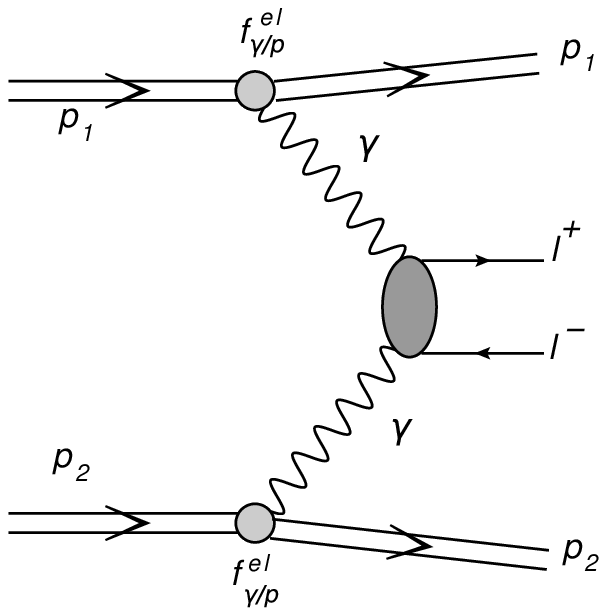}
\includegraphics[width=5cm]{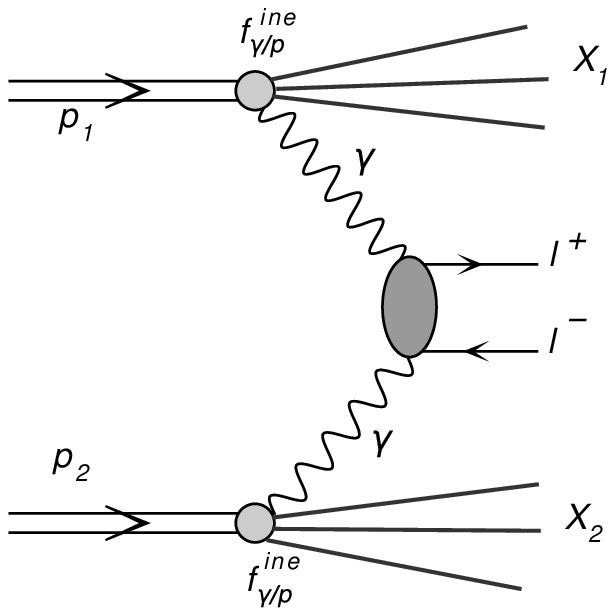}\\
\includegraphics[width=5cm]{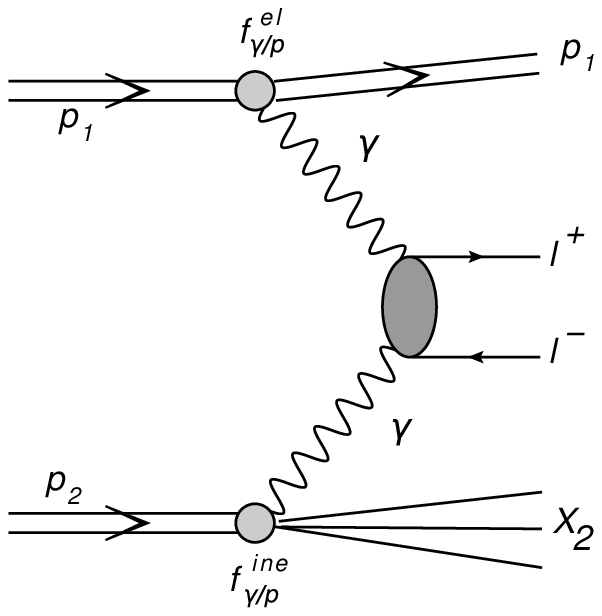}
\includegraphics[width=5cm]{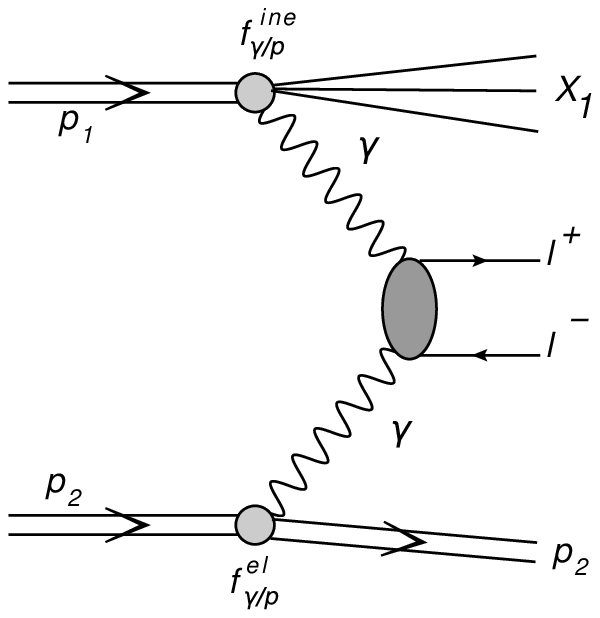}
\end{center}
\caption{
Different mechanisms of two-photon production of dileptons.
}
\label{fig:diagrams}
\end{figure}

In this paper we wish to concentrate on the photon-photon
induced production of charged leptons.
Realistic estimation of these processes requires more attention.
In general, there are three types of such processes which
can be classified according to whether the proton remnants appearing
``after'' photon emission are just protons or baryon resonances 
or a complicated continuum (see Fig.\ref{fig:diagrams}).
In principle, the elastic-elastic, processes with one elastic and one
inelastic or double inelastic processes can be distinguished by
detailed studies of the final state. However, in practice this
separation may be not easy and all of them should be considered.
Here we wish to concentrate rather on inelastic-inelastic processes.

There are two approaches in the literature in this context.
In one of the approaches one can treat photons as collinear partons in the proton.
The application of this approach requires presence of a hard scale
(e.g. a large photon virtuality or large lepton transverse momenta).
Such photon partonic distributions were discussed in 
\cite{Gluck:2002fi,Martin:2004dh,Ball:2013hta,Martin:2014nqa,Schmidt:2014aba}. In 
some of these approaches the photon PDF enters the
DGLAP evolution equations.
The treatment in \cite{Gluck:2002fi} is somewhat simplified, 
here only the $q \to \gamma$ splitting is taken into account. 

Below we wish to comment on the interrelation between
the two approaches. The photon PDF approach was applied in
many phenomenological studies, e.g. to a number of photon-photon 
processes in \cite{Luszczak:2014mta}, and to 
dilepton production in \cite{Maciula:2010yw}. 

In another approach one parametrizes the $\gamma^* p \to X$ vertices
in terms of the proton's structure functions. One can assume that 
the photons are either collinear or allow them to have 
transverse momenta and nonzero virtualities \cite{Budnev:1974de}.
Recently we used a slightly simplified approach \cite{daSilveira:2014jla},
which takes advantage of the high-energy limit and is 
formulated in an analogous way as the $k_T$-factorization approach often used 
in the context of two-gluon processes. In this approach one uses
unintegrated photon distributions, in contrast to collinear
distributions in the previous approach and off-shell matrix elements
for the $\gamma \gamma \to l^+ l^-$ subprocess.
We shall use this approach also in the present paper.

The unintegrated photon distributions can be expressed in terms
of the proton structure functions. The structure functions
were measured in some different corners of very rich phase space. 
In particular they were studied in so-called deep-inelastic regime
with large $Q^2$ where perturbative treatment embedded in the DGLAP
evolution equation applies. In this corner of the phase space
the structure functions are very well known.
When going outside of the perturbative regime the situation is less
clear. Several parametrizations were presented in the literature
\cite{Suri:1971yx,Fiore:2002re,Szczurek:1999rd,Block:2014kza, 
Abramowicz:1991xz,Abramowicz:1997ms}. 
The applicability of the different
parametrizations is limited and not well tested.

Thus in two-photon processes one may need structure functions
in very different corners of the $(x, Q^2)$ space.
It is not clear a priori which regions are needed for particular
experiments i.e. specific kinematical cuts. We wish to discuss
some examples related to particular past and modern experiments.


This paper is organized as follows: in section \ref{sec:fluxes}, we 
briefly review the different formalism employed in our calculations.
We also discuss the different structure functions used as an input
in the $k_T$-factorization approach.
In section \ref{sec:results} we show our numerical results of various
dilepton distributions for the kinematics, and cuts, relevant for
different experiments. These are, at the presently highest available energies,
ATLAS and CMS, which measure central rapidities, and LHCb with coverage 
at forward rapidities. We also discuss examples for the lower energies of RHIC,
as well as data taken in 1980's at the ISR at still lower energy.
We summarize our results in Conclusions section.  


\section{Collinear-factorization approach}

\subsection{Photons as partons in a hard process}
\label{subsec:DGLAP-photons}

Production of lepton pairs at large transverse momenta is a hard process, to which
standard arguments for factorization apply, and collinear factorization should be an appropriate 
starting point to calculate e.g. rapidity or transverse momentum spectra of leptons.
In fact, the dominant contribution to large-invariant mass dilepton pairs is of course
the well known Drell-Yan process, but nothing prevents us from also including photon
as partons along with quarks and gluons.

Then the photon parton distribution, $\gamma(z,Q^2)$, of photons carrying a fraction $z$ of the proton's
light-cone momentum, obeys the DGLAP equation,
\begin{eqnarray}
{d \gamma(z,Q^2) \over d \log Q^2} =&& {\alpha_{\rm{em}} \over 2 \pi} \int_x^1 {dy \over y} 
\Big \{ \sum_f e_f^2 P_{\gamma \leftarrow q}(y) 
\Big[ q_f \Big({z \over y}, Q^2 \Big) + \bar q_f\Big({z \over y},Q^2\Big) \Big] \nonumber \\
&&+ P_{\gamma \leftarrow \gamma}(y) \gamma\Big({z \over y},Q^2\Big) \Big \} \, .
\end{eqnarray}
In the complete set of DGLAP equations this photon density is then again coupled to the quark and antiquark
distributions:
\begin{eqnarray}
 {d q_f(z,Q^2 )\over d \log Q^2} =&& {d q_f(z,Q^2) \over d \log Q^2}\Big|_{\rm{QCD}} + {\aem \over 2 \pi} \int_x^1 {dy \over y} 
\delta P_{q \leftarrow q}^{\rm{QED}}(y) q_f\Big({z \over y}, Q^2\Big)
\nonumber \\
+&& {\aem \over 2 \pi} \int_x^1 {dy \over y}  P_{q \leftarrow \gamma}(y)
\gamma\Big({z \over y},Q^2\Big) \; . \nonumber \\
\end{eqnarray}
Due to the smallness of $\alpha_{\rm{em}}$ one would expect that the effect of photons on the quark and antiquark densities
can be safely neglected, unless one is interested in high order perturbative corrections to the QCD splitting functions
themselves.

Accordingly, we find two different approaches to DGLAP photons in the literature.

A first one, by Gl\"uck et al. \cite{Gluck:2002fi} asserts, that 
we can neglect the photon density on the right hand side of the evolution equations.
Then, at sufficiently large virtuality $Q_0^2$, the photon parton density can
be calculated from the collinear splitting of quarks and antiquarks 
$q \to q \gamma, \bar q \to \bar q  \gamma$. 
\begin{eqnarray}
 {d \gamma(z,Q^2)\over d \log Q^2} = {\aem \over 2 \pi} \sum_f  e_f^2 \int_z^1 {dx \over x}   P_{\gamma \leftarrow q}\Big({z \over x}\Big)
\Big[ q_f(x,Q^2) + \bar q_f(x,Q^2) \Big] \; .
\label{eq:Dortmund}
\end{eqnarray}
This equation is easily integrated, and gives the photon parton density as
\begin{eqnarray}
 \gamma(z,Q^2) &=& \sum_f { \aem e_f^2 \over 2 \pi} \int_{Q_0^2}^{Q^2} {d \mu^2 \over \mu^2} 
\int_z^1 {dx \over x}  P_{\gamma \leftarrow q}\Big({z \over x}\Big)
\Big[ q_f(x,\mu^2) + \bar q_f(x,\mu^2) \Big] + \gamma(z,Q_0^2) \nonumber \\
&=& {\aem \over 2 \pi} \int_{Q_0^2}^{Q^2} {d \mu^2 \over \mu^2} \int_z^1 {dx \over x}  
P_{\gamma \leftarrow q}\Big({z \over x}\Big) {F_2(x,\mu^2) \over x} 
+ \gamma(z,Q_0^2) \, .
\end{eqnarray}
One is left to specify -- from some  model considerations -- the photon density at some low scale $\gamma(z,Q_0^2)$, but
one may hope that at very large $Q^2 \gg Q_0^2 \sim 1 \, {\rm GeV}^2$ the part predicted perturbatively 
from quark and antiquark distributions dominates.

In addition to the above contribution from DGLAP splitting, Gl\"uck et
al. also add the Weizs\"acker-Williams flux from the coherent emission 
$p \to p \gamma^*$ without proton breakup as found in \cite{Budnev:1974de}.
 
More recently, the Durham \cite{Martin:2004dh,Martin:2014nqa} and NNPDF 
\cite{Ball:2013hta} groups have given a more involved treatment, 
in which the photon distribution is fully incorporated into the coupled 
DGLAP evolution equation.
As usual with DGLAP evolution, the photon parton density at a starting scale 
$\gamma(z,Q_0^2)$ needs to be specified. While \cite{Martin:2004dh,Martin:2014nqa} present
model approaches, in Ref.\cite{Ball:2013hta} an ambitious attempt to
obtain $\gamma(z,Q_0^2)$ from a fit to experimental data is found.
Preliminary work by the CTEQ collaboration \cite{Schmidt:2014aba} is
also based on QED corrected DGLAP equations,
and attempts to fit the photon distribution from the prompt photon
production $e p \to \gamma e X$ at HERA
where in part of the phase space the Compton subprocess $e \gamma \to e \gamma$ contributes.

It should be noted, that in the approach of
\cite{Martin:2004dh,Martin:2014nqa}, the input distribution $\gamma(z,Q_0^2)$ contains
the coherent --or elastic-- contribution with an intact proton in the final state.
Notice that due to the proton form factors the integral over
virtualities in the elastic case quickly converges, and the elastic
contribution is basically independent of $Q_0^2$, as soon as $Q_0^2 \gsim 0.7 \, \rm{GeV}^2$.

\subsection{From photon PDFs to cross section}

In the collinear approach the photon-photon contribution
to inclusive cross section for dilepton production can be written as:
\begin{equation}
{d \sigma^{(i,j)} \over d y_1 d y_2 d^2 p_T} 
= {1 \over 16 \pi^2 (x_1 x_2 s)^2}\sum_{i,j} 
x_1 \gamma^{(i)}(x_1,\mu^2) 
x_2 \gamma^{(j)}(x_2,\mu^2)
\overline{ |{\cal M}_{\gamma \gamma \rightarrow l^+ l^-}|^2 }.
\label{collinear_factorization_formula}
\end{equation}
Here 
\begin{eqnarray}
 x_1 &=&  \sqrt{p_T^2 + m_l^2 \over s} 
\Big( \exp(y_1) + \exp(y_2) \Big) \; , \nonumber \\
 x_2 &=&  \sqrt{p_T^2 + m_l^2 \over s} 
\Big( \exp(-y_1) + \exp(-y_2) \Big) \; . 
\end{eqnarray}
Above indices $i$ and $j$ denote $i,j = \rm{el, in}$, i.e. they
correspond to elastic or inelastic components similarly as for 
the $k_T$-factorization discussed in section \ref{sec:fluxes} below, 
see also the diagrams in Fig.\ref{fig:diagrams}.
The factorization scale is chosen as $\mu^2 = m_T^2 = p_T^2 + m_l^2$.

The elastic photon distributions can be calculated with the help
of elastic electromagnetic proton form factors.
The functions $\gamma^{({\rm{in}})}(x,\mu^2)$ are precisely the DGLAP evolved
distributions of section \ref{subsec:DGLAP-photons} above.

\subsection{Initial condition for collinear photon PDF}

In the MRST2004(QED) approach the initial photon distribution is
parametrized as \cite{Martin:2014nqa}:
\begin{equation}
\gamma(z,Q_0^2) =  {\aem \over 2 \pi} \int_z^1 {dy \over y}
\left[ \frac{4}{9} \log\left(\frac{Q_0^2}{m_u^2}\right) u\Big({z \over y},Q_0^2 \Big)
 + \frac{1}{9} \log\left(\frac{Q_0^2}{m_d^2}\right) d\Big({z \over y},Q_0^2\Big) \right]
\cdot \frac{1+(1-y)^2}{y} \;.
\label{photon_initial}
\end{equation}
Above $u(x,Q_0^2)$ and $d(x,Q_0^2)$ are valence-like distributions
at the initial scale $Q_0^2$.
In actual calculation MRST2004(QED) uses current quark masses which 
causes that the $\log(\frac{Q_0^2}{m_q^2})$'s and in the consequence also
the initial photon distributions are artificially large
(the consequences for lepton production will be discussed 
when showing corresponding cross sections).
It would seem more reasonable to use rather 
constituent quark masses than the current ones.
We will show that this leads to large differences in photon
distributions at finite running scales $Q^2$. 

\begin{figure}
\begin{center}
\includegraphics[width=8cm]{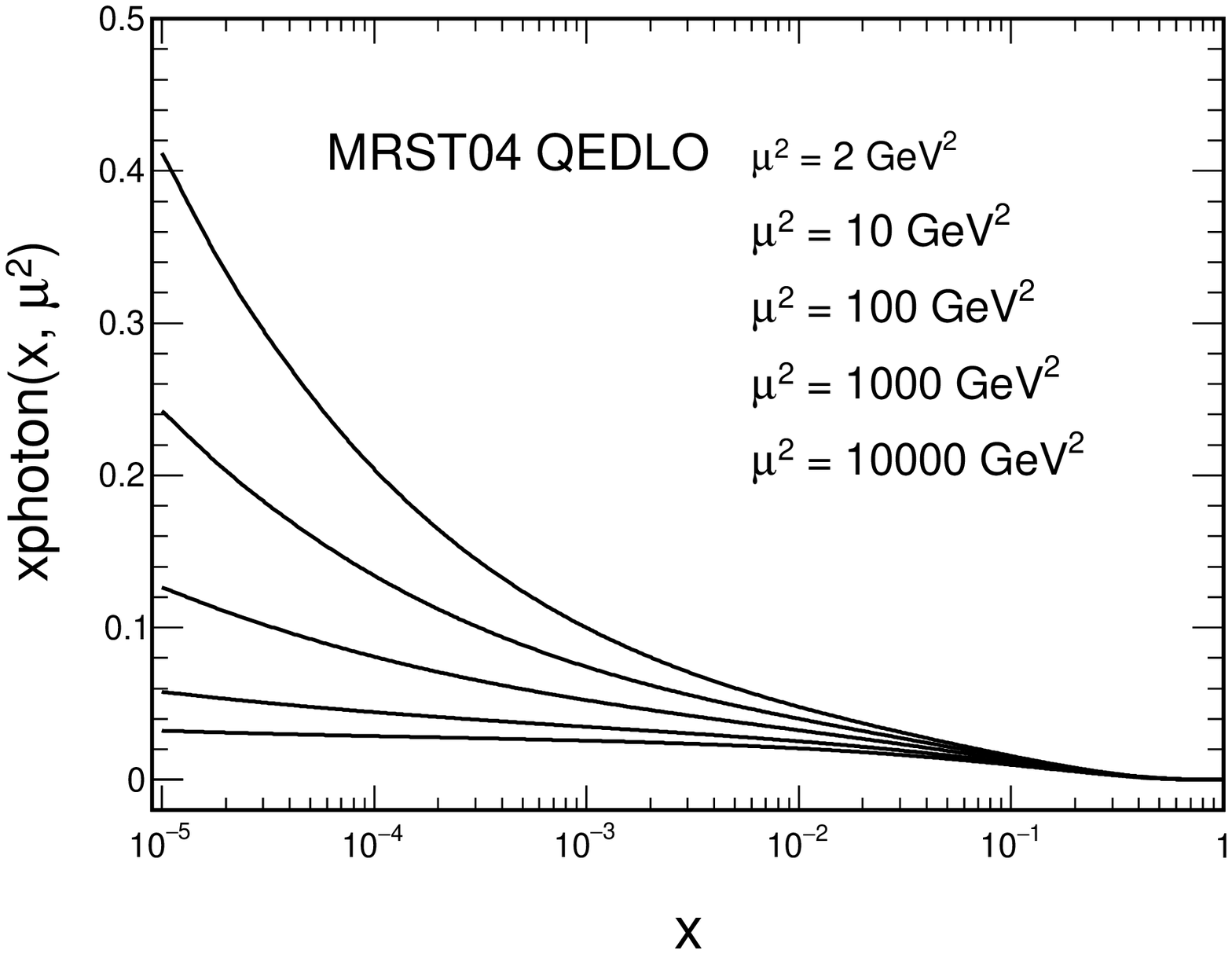}
\includegraphics[width=8cm]{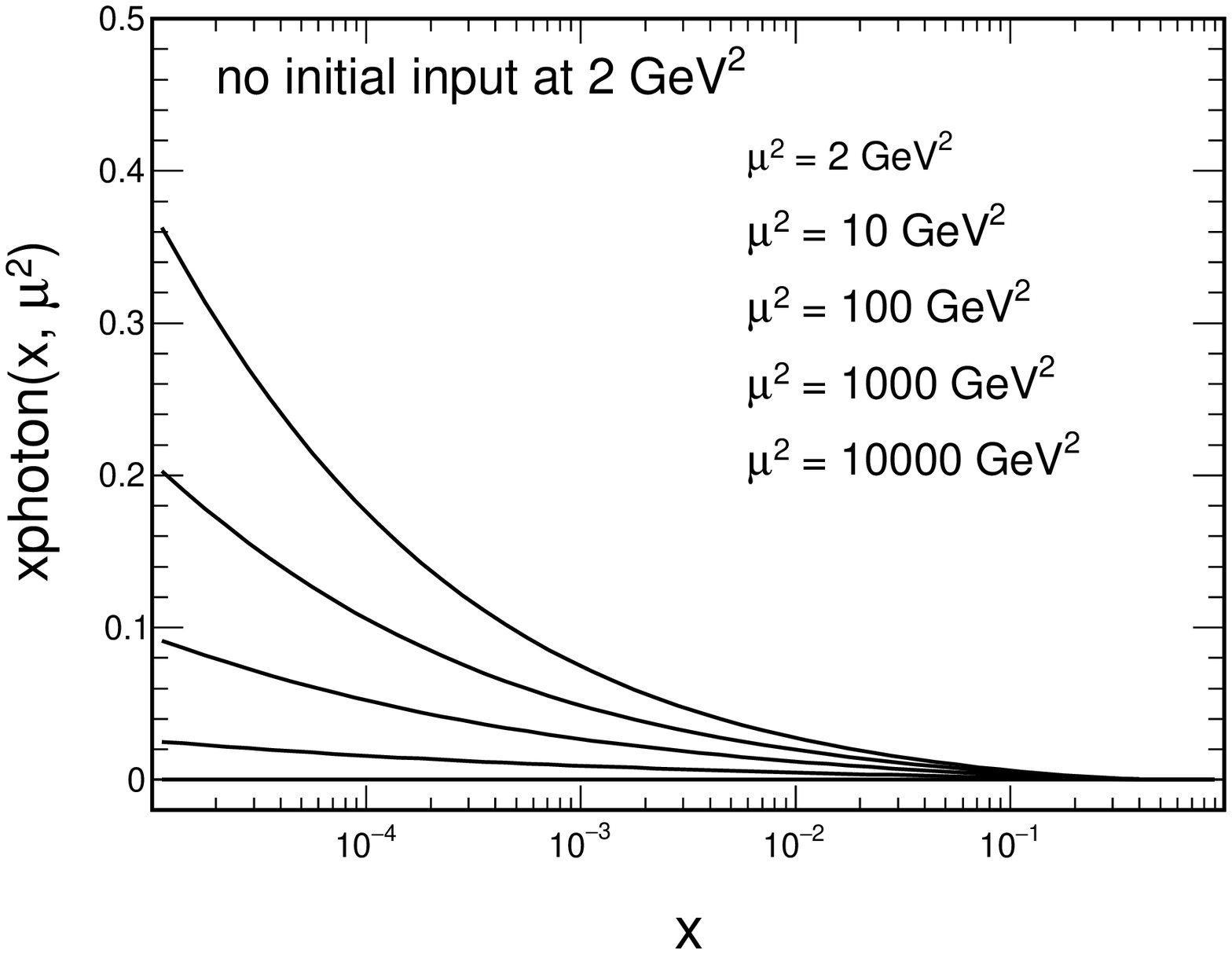}
\end{center}
\caption{
Collinear photon distributions for different scales.
The left panel is for standard MRST2004(QED) parton distribution,
while the right panel is for the case when the initial input
at $\mu^2$ = 2 GeV$^2$ is set to zero, i.e. completely neglected.
}
\label{fig:collinear_photon_pdf}
\end{figure}

Before discussing results for cross sections for $l^+ l^-$ production
we wish to concentrate for a while on the collinear photon distributions.
To illustrate the effect of the initial input in 
Fig.\ref{fig:collinear_photon_pdf} we show both original
MRST2004(QED) photon distribution and similar result
obtained by ignoring the initial input which, as discussed above,
may be questionable.
The results are shown for different evolution scales 
$\mu^2 = Q_0^2$, 10, 100, 1000, 10000 GeV$^2$.
We observe a sizable difference between resulting photon distributions 
obtained within the two approaches.
Because in calculating the cross section the photon distributions
enter twice in the cross section formula, for first and second 
proton, respectively, 
one can expect that the cross section obtained with the different
PDFs may differ considerably.
We will return to this issue in the Result section.

\section{$k_T$-factorization approach}
\label{sec:fluxes}

In this approach we start from the Feynman diagrams shown in Fig.\ref{fig:diagrams},
and exploit the high-energy kinematics.
Let the four-momenta of incoming protons be denoted $p_A,p_B$. At high energies 
the proton masses can be neglected, 
so that $p_A^2 = p_B^2 =0, \,  2 (p_A\cdot p_B) =s$.

The photon-fusion production mechanism in leptonic and hadronic reactions
is in great detail reviewed in \cite{Budnev:1974de}, where also many original
references can be found. In the most general form, the invariant cross section
is written as a convolution of density matrices of photons in the beam particles,
and helicity amplitudes for the $\gamma^* \gamma^* \to l^+ l^-$ process.
In a high energy limit, where dileptons carry only a small fraction of the 
total center-of-mass energy, the density-matrix structure can be very much
simplified, and there emerges a $k_T$-factorization representation of the
cross section \cite{daSilveira:2014jla}.

The unintegrated photon fluxes introduced in \cite{daSilveira:2014jla}
can be expressed in terms of the hadronic tensor as 
\begin{eqnarray}
 {\cal{F}}^{{\rm{in.el}}}_{\gamma^* \leftarrow A} (z,\bq) = {\alpha_{\rm{em}}\over \pi}  \, (1-z) \, 
\Big( {\bq^2 \over \bq^2 + z (M_X^2 - m_A^2) + z^2 m_A^2  }\Big)^2  \, 
\cdot {p_B^\mu p_B^\nu \over s^2} \, W^{\rm{in,el}}_{\mu \nu}(M_X^2,Q^2) dM_X^2 \, . \nonumber \\
\end{eqnarray}
These unintegrated fluxes enter the cross section for dilepton production as
\begin{eqnarray}
 {d \sigma^{(i,j)} \over dy_1 dy_2 d^2\bp_1 d^2\bp_2} &&=  \int  {d^2 \bq_1 \over \pi \bq_1^2} {d^2 \bq_2 \over \pi \bq_2^2}  
 {\cal{F}}^{(i)}_{\gamma^*/A}(x_1,\bq_1) \, {\cal{F}}^{(j)}_{\gamma^*/B}(x_2,\bq_2) 
{d \sigma^*(p_1,p_2;\bq_1,\bq_2) \over dy_1 dy_2 d^2\bp_1 d^2\bp_2} \, , \nonumber \\ 
\label{eq:kt-fact}
\end{eqnarray}
where the indices $i,j \in \{\rm{el}, \rm{in} \}$ denote elastic or inelastic final states.
The longitudinal momentum fractions of photons are obtained from the rapidities 
and transverse momenta of final state leptons as:
\begin{eqnarray}
x_1 &=& \sqrt{ {\bp_1^2 + m_l^2 \over s}} e^{y_1} +  \sqrt{ {\bp_2^2 +
    m_l^2 \over s}} e^{y_2} 
\; , \nonumber \\
x_2 &=& \sqrt{ {\bp_1^2 + m_l^2 \over s}} e^{-y_1} +  \sqrt{ {\bp_2^2 + m_l^2 \over s}} e^{-y_2} \, .
\end{eqnarray}
The explicit form of the off-shell cross section $d \sigma^*(p_1,p_2;\bq_1,\bq_2)/ dy_1 dy_2 d^2\bp_1 d^2\bp_2$ can be found in
Ref. \cite{daSilveira:2014jla}. 

\subsection{Inelastic vertices}

We now first concentrate on inelastic processes with breakup of a proton. 
Then the hadronic tensor is expressed in terms of the electromagnetic currents as:
\begin{eqnarray}
 W^{\rm{in}}_{\mu \nu}(M_X^2,Q^2) = \overline{\sum_X} (2 \pi)^3 \, \delta^{(4)} (p_X - p_A - q) \, \bra{p} J_\mu \ket{X}\bra{X} J_\nu^\dagger \ket{p} \, d\Phi_X \, ,
\label{eq:Wmunu}
\end{eqnarray}
and its elements can be measured in inclusive electron scattering 
off the target. We wish to express it in terms of the virtual photoabsorption cross section
of transverse and longitudinal photons. To this end we introduce the covariant vectors/tensors
\begin{eqnarray}
e_\mu^{(0)} = \sqrt{Q^2 \over  X} \Big( p_{A\mu} - { (p_A \cdot q ) \over q^2} q_\mu \Big) \, , \, 
X = (p_A \cdot q)^2 + m_A^2 Q^2 \, , \, e^{(0)}\cdot e^{(0)} = + 1 \, ,
\end{eqnarray}
and
\begin{eqnarray}
\delta^\perp_{\mu \nu}(p_A,q) = g_{\mu \nu} - {q_\mu q_\nu \over q^2} - e^{(0)}_\mu e^{(0)}_\nu \, .
\end{eqnarray}
Here $\delta^\perp_{\mu\nu}$ projects on photons carrying helicity $\pm 1$ in the $\gamma^* p$-cms frame,
and $e_\mu^{(0)}$ plays the role of the polarization vector of the longitudinal photon.
Notice that $q\cdot e^{0} = q^\mu \delta^\perp_{\mu \nu} = 0$, so that the hadronic tensor has the convenient
gauge invariant decomposition
\begin{eqnarray}
  W^{\rm{in}}_{\mu \nu}(M_X^2,Q^2) = - \delta^\perp_{\mu \nu} (p_A,q) \, W^{\rm{in}}_T(M_X^2, Q^2) + e^{(0)}_\mu e^{(0)}_\nu \, W^{\rm{in}}_L(M_X^2, Q^2) \, .
\end{eqnarray}
The virtual photoabsorption cross sections are defined as
\begin{eqnarray}
 \sigma_T(\gamma^* p) &=& {4 \pi \aem \over 4 \sqrt{X}} \, \Big(- {\delta^\perp_{\mu\nu} \over 2} \Big)  2\pi W^{\rm{in}}_{\mu \nu}(M_X^2,Q^2) \nonumber \\
 \sigma_L(\gamma^* p) &=& {4 \pi \aem \over 4 \sqrt{X}} \, e^{0}_\mu e^{0}_\nu \, 2 \pi W^{\rm{in}}_{\mu \nu}(M_X^2,Q^2) \, .
\end{eqnarray}
It is customary to introduce dimensionless structure function $F_i(x_{\rm Bj},Q^2), i = T,L$ as
\begin{eqnarray}
 \sigma_{T,L}(\gamma^* p) = {4 \pi^2 \aem \over Q^2} \, {1 \over \sqrt{1 + {4 x^2_{\rm Bj} m_A^2 \over Q^2}} } \, F_{T,L}(x_{\rm Bj},Q^2) \, ,
\end{eqnarray}
where
\begin{eqnarray}
 x_{\rm Bj} = { Q^2 \over Q^2 + M_X^2 - m_A^2} \, .
\end{eqnarray}
Then, our structure functions $W_{T,L}$ are expressed through the more conventional $F_{T,L}$ as
\begin{eqnarray}
 W^{\rm{in}}_{T,L}(M_X^2,Q^2) = {1 \over x_{\rm Bj}} \, F_{T,L}(x_{\rm Bj},Q^2) \, . 
\end{eqnarray}
In the literature one often finds rather $F_1(x_{\rm Bj}, Q^2), F_2(x_{\rm Bj},Q^2)$
structure functions, which are related to $F_{T,L}$ through
\begin{eqnarray}
 F_T(x_{\rm Bj},Q^2) &=& 2x_{\rm Bj}  F_1(x_{\rm Bj},Q^2) \, , \nonumber \\
F_2(x_{\rm Bj},Q^2)  &=& { F_T(x_{\rm Bj},Q^2) +F_L(x_{\rm Bj},Q^2)
  \over 1 + {4 x^2_{\rm Bj} m_A^2 \over Q^2}} \; .
\end{eqnarray}
Now, performing the contraction with $p^\mu_B p^\nu_B$, we get
\begin{eqnarray}
  {p_B^\mu p_B^\nu \over s^2} \, W^{\rm in}_{\mu \nu}(M_X^2,Q^2) = \Big( 1 - {z \over x_{\rm Bj}} + {z^2\over 4 x_{\rm Bj}^2} \Big) {F_2(x_{\rm Bj},Q^2) \over Q^2 + M_X^2 - m_p^2} 
+ {z^2 \over 4 x^2_{\rm Bj}} { 2 x_{\rm Bj} F_1(x_{\rm Bj},Q^2) \over Q^2 + M_X^2 - m_p^2} \, .
\end{eqnarray}
In the deep inelastic region $F_2 \sim F_T + F_L$, and using $2x_{\rm Bj} F_1 \sim F_2$ in the second term, we can write more succinctly
\begin{eqnarray}
  {p_B^\mu p_B^\nu \over s^2} \, W^{\rm in}_{\mu \nu}(M_X^2,Q^2) = Q^2 \cdot f_T \Big( {z \over x_{\rm Bj}} \Big) \,  x_{\rm Bj} F_2(x_{\rm Bj},Q^2) \, ,
\end{eqnarray}
with 
\begin{eqnarray}
 f_T(y) = 1 - y + y^2/2 = {1 \over 2} \Big[ 1 + (1-y)^2 \Big] \, .
\end{eqnarray}

\subsection{Elastic vertices}

Let us now isolate the elastic contribution to the hadronic tensor,
which we need to describe the photon flux in processes in which the
proton stays intact. In this case, the structure functions $W_{T,L}$ are most 
conveniently written in terms of the electric and magnetic form
factors $G_E(Q^2)$ and $G_M(Q^2)$ of the proton:
\begin{eqnarray}
W^{\rm el}_T(M_X^2,Q^2) = \delta(M_X^2 - m_p^2) \, Q^2 G^2_M(Q^2) \, , \, W^{\rm el}_L(M_X^2, Q^2) = \delta(M_X^2 - m_p^2) 4m_p^2 G^2_E(Q^2) \, . 
\end{eqnarray}
The contribution to the photon flux is then again obtained by contracting
\begin{eqnarray}
  {p_B^\mu p_B^\nu \over s^2} \, W^{\rm el}_{\mu \nu}(M_X^2,Q^2) = \delta(M_X^2 - m_p^2) 
\Big[ \Big( 1- {z \over 2} \Big)^2 \, {4 m_p^2 G_E^2(Q^2) + Q^2 G_M^2(Q^2) \over 4m_p^2 + Q^2} + {z^2 \over 4} G_M^2(Q^2) \Big]
\nonumber \\
\end{eqnarray}

\subsection{Unintegrated photon fluxes}

Let us now give explicit formulas for the unintegrated fluxes in a form which makes it easy to compare them
for example with fluxes of virtual photons given by Budnev et al. \cite{Budnev:1974de}.
The quantity to compare is the differential equivalent photon spectrum
\begin{eqnarray}
 dn^{\mathrm{in,el}} = {dz \over z} {d^2 \bq \over \pi \bq^2} \, {\cal{F}}^{\mathrm{in,el}}_{\gamma^* \leftarrow A} (z,\bq) \, .
\end{eqnarray}
The fluxes in \cite{Budnev:1974de} are given differentially in the virtuality $Q^2$, instead
of the transverse momentum $\bq^2 = (1-z) (Q^2 - Q^2_{\rm min})$. We therefore substitute
\begin{eqnarray}
  {d^2 \bq \over \pi \bq^2} \to (1-z) {dQ^2 \over Q^2} \cdot {Q^2 \over \bq^2} = {dQ^2 \over Q^2} \cdot {Q^2 \over Q^2 - Q^2_{\rm min}} \, 
{\rm{and}} \, { \bq^2 \over \bq^2 + z (M_X^2 - m_A^2) + z^2 m_A^2 }= {Q^2 - Q^2_{\rm min} \over Q^2} \, , \nonumber \\
\end{eqnarray}
so that we obtain
\begin{eqnarray}
 dn^{\rm in} &=& {\aem \over \pi} {dQ^2 \over Q^2} {dz \over z} (1-z) \Big( 1 - {Q^2_{\rm min} \over Q^2} \Big) \, 
\nonumber \\
&\times& \Big[\Big( 1 - {z \over x_{\rm Bj}} + {z^2\over 4 x_{\rm Bj}^2} \Big) {F_2(x_{\rm Bj},Q^2) \over Q^2 + M_X^2 - m_p^2} 
+ {z^2 \over 4 x^2_{\rm Bj}} { 2 x_{\rm Bj} F_1(x_{\rm Bj},Q^2) \over
  Q^2 + M_X^2 - m_p^2} \Big] 
\; dM_X^2
\end{eqnarray}
and for the elastic piece
\begin{eqnarray}
 dn^{\rm el} = {\aem \over \pi} {dQ^2 \over Q^2} {dz \over z} (1-z) \Big( 1 - {Q^2_{\rm min} \over Q^2} \Big) \,
\Big[ \Big( 1- {z \over 2} \Big)^2 \, {4 m_p^2 G_E^2(Q^2) + Q^2
  G_M^2(Q^2) \over 4m_p^2 + Q^2} + {z^2 \over 4} G_M^2(Q^2) \Big] \; .
\nonumber \\
\end{eqnarray}
It is also interesting to convert the integration over $M_X^2$ into one over $x_{\rm Bj}$.
To this end, we note that 
\begin{eqnarray}
 {dM_X^2 \over Q^2 + M_X^2 - m_p^2} \to {dx_{\rm Bj} \over x_{\rm Bj}} \, , \, x_{\rm min} = {z \over 1 - z^2 {m_p^2 \over Q^2}} \, , \,  
x_{\rm max} = {Q^2 \over Q^2 + (2 m_p + m_\pi)m_\pi} \, .
\end{eqnarray}
Furthermore
\begin{eqnarray}
 (1-z) \Big( 1 - {Q^2_{\rm min} \over Q^2} \Big) = {z \over x_{\rm Bj} } \, \Big( {x_{\rm Bj} \over x_{\rm min}} -1 \Big) =  1 - {z \over x_{\rm Bj} }- {z^2 m_p^2 \over Q^2} \, .
\end{eqnarray}
Then we obtain for the photon flux
\begin{eqnarray}
 {z dn^{\rm{in}}(z,Q^2) \over dz d\log Q^2 } = {\aem \over \pi} \int_{x_{\rm min}}^{x_{\rm max}}  {dx_{\rm Bj} \over x_{\rm Bj}} 
&&\Big( 1 - {z \over x_{\rm Bj} }- {z^2 m_p^2 \over Q^2} \Big)
\Big[\Big( 1 - {z \over x_{\rm Bj}} + {z^2\over 4 x_{\rm Bj}^2} \Big) F_2(x_{\rm Bj},Q^2) 
\nonumber \\
&& + {z^2 \over 4 x^2_{\rm Bj}}  2 x_{\rm Bj} F_1(x_{\rm Bj},Q^2) \Big]  .
\end{eqnarray}
In the deep inelastic limit, $x_{\rm{min}} \to z, x_{\rm{max}} \to 1$, and assuming $F_2 = 2 x_{\rm Bj} F_1$, this obtains the form 
\begin{eqnarray}
 {dn^{\rm{in}}(z,Q^2) \over dz d\log Q^2} = {\aem \over 2 \pi} \int_z^1  {dx_{\rm Bj} \over x_{\rm Bj}} \, P_{\gamma \leftarrow q}\Big({z \over x_{\rm Bj}} \Big) \, { F_2(x_{\rm Bj},Q^2) \over x_{\rm Bj}} 
\, \Big( 1 - {z \over x_{\rm Bj} } \Big) \, ,
\end{eqnarray}
with the splitting function
\begin{eqnarray}
  P_{\gamma \leftarrow q}(y) = {1 + (1-y)^2 \over y} \, .
\end{eqnarray}
The ``parton densities of photons'', which can be compared to the collinear factorization fluxes 
are
\begin{eqnarray}
\gamma^{\rm in,el} (z,\mu^2) = \int^{\mu^2} {dQ^2 \over Q^2}   {dn^{\rm{in,el}}(z,Q^2) \over dz d\log Q^2} \, .
\end{eqnarray}
%
\subsection{Structure functions as input for unintegrated fluxes}

Here we show a few different parametrizations of the proton structure function $F_2$. 

The different parametrizations taken from the literature are labeled as:
\begin{itemize} 
 \item ALLM \cite{Abramowicz:1991xz,Abramowicz:1997ms}. This
   parametrization gives a very good fit to $F_2$ in most of the measured region.
\item FJLLM \cite{Fiore:2002re}. This parametrization explicitly
  includes the nucleon resonances and gives an excellent fit of the CLAS data.
 \item BDH \cite{Block:2014kza}. This parametrization concentrates on
   the low-$x$, or high mass region. It features a Froissart-like behaviour at very small $x$. 
  \item SY \cite{Suri:1971yx}. This paramerization of Suri and Yennie
    from the early 1970's does not include QCD-DGLAP evolution. It is
    still today often used as one of the defaults in the LPAIR event generator.
 \item SU \cite{Szczurek:1999rd}. A parametrization which concentrates
   to give a good description at smallish and intermediate $Q^2$ at not too small $x$.
\end{itemize}

We also show $F_2$ calculated from the CTEQ6L parametrization \cite{Pumplin:2002vw}.

\begin{figure}[!h]
   \includegraphics[width = 0.4\textwidth]{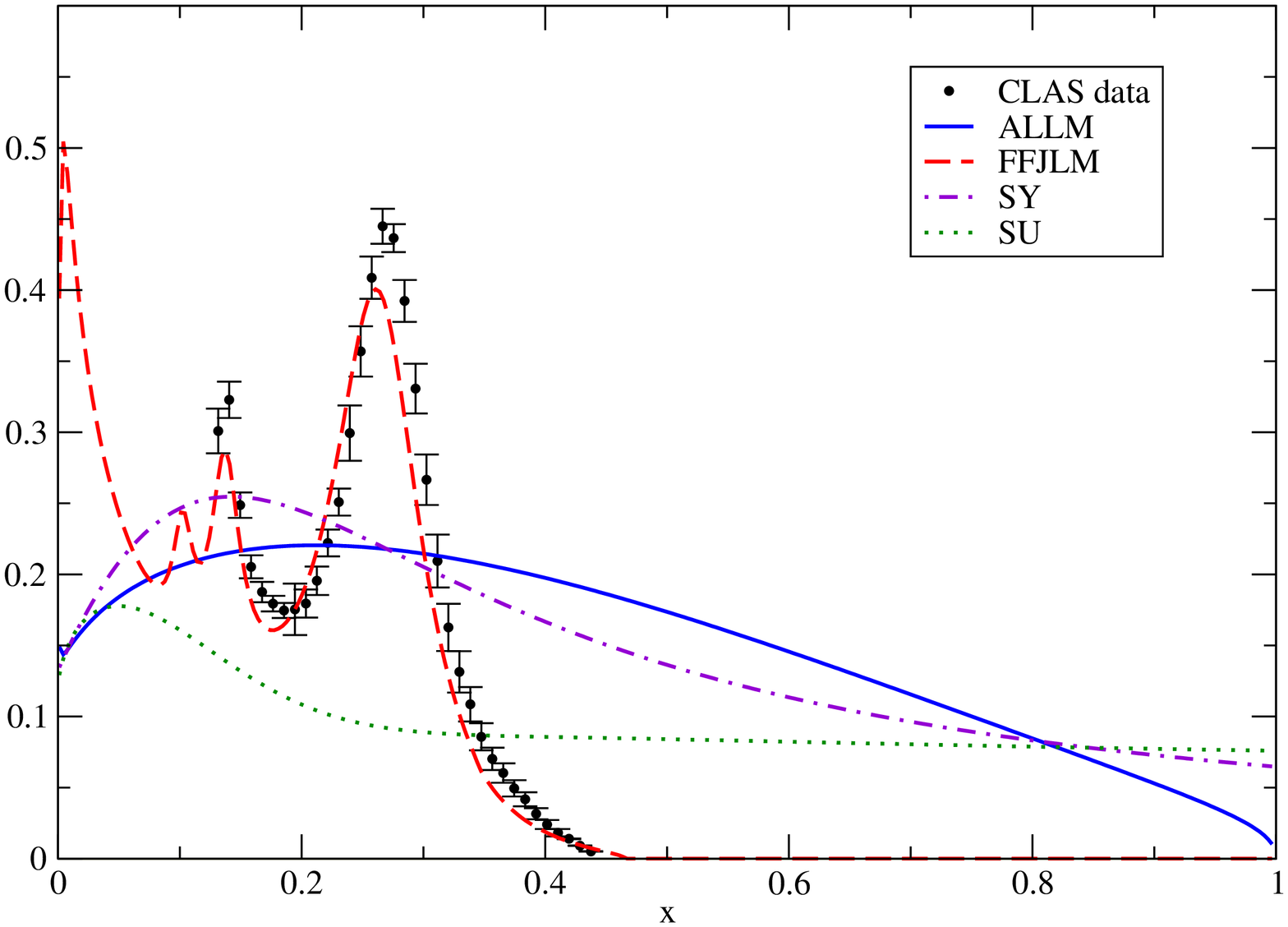}
   \includegraphics[width = .4 \textwidth]{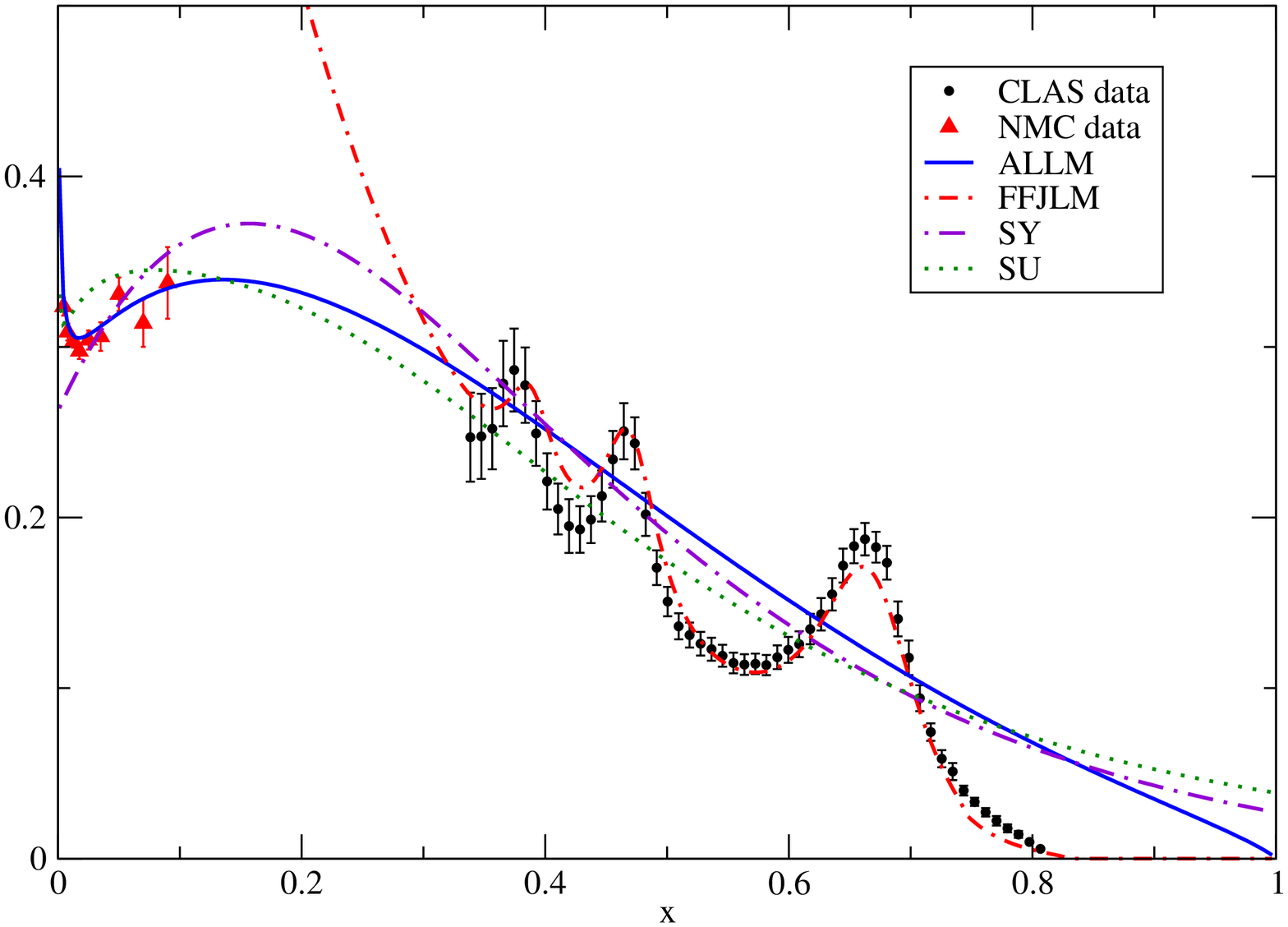}
   \includegraphics[width=.4 \textwidth]{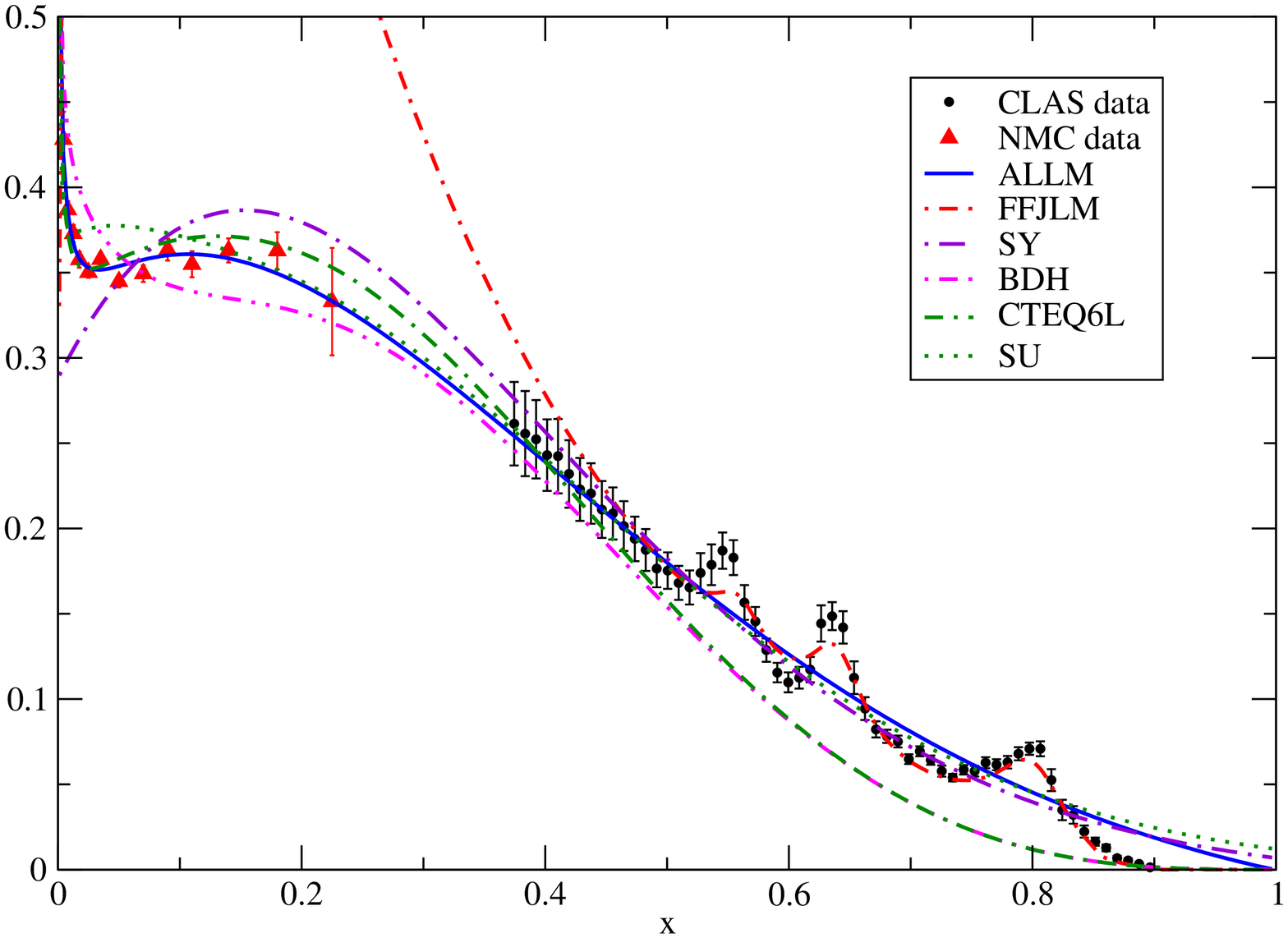}
   \includegraphics[width = 0.4\textwidth]{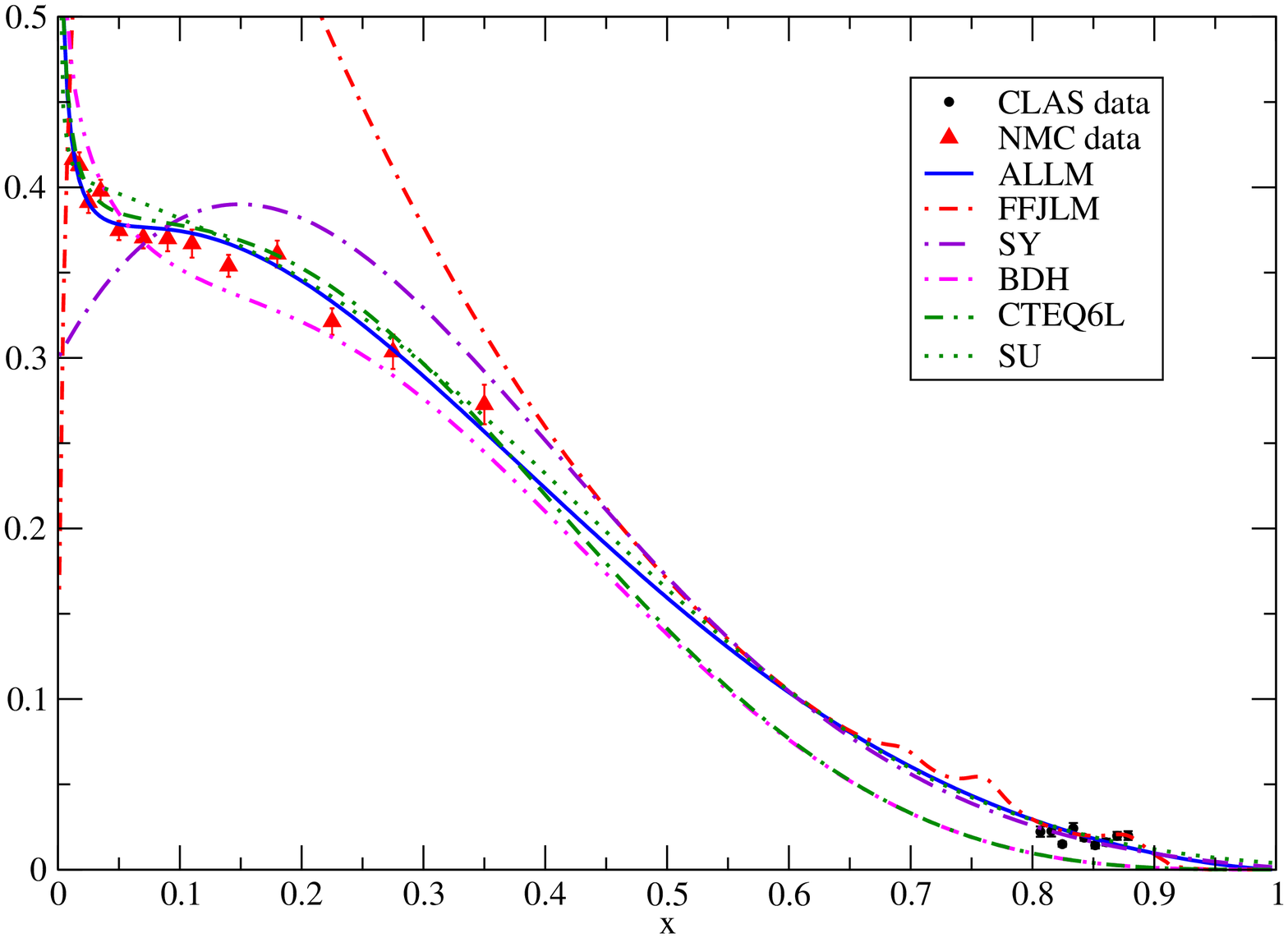}
   \caption{The proton structure function $F_2(x,Q^2)$ as a function of $x$ for $Q^2= 0.225 \, \rm{GeV}^2$(top left),
$Q^2 = 1.25 \, \rm{GeV}^2$(top right), $Q^2 = 2.5 \, \rm{GeV}^2$(bottom left), and $Q^2= 4.5 \, \rm{GeV}^2$(bottom right).
Shown are different parmetrizations available in the literature.
 }
  \label{fig:F2}
 \end{figure}

\begin{figure}[!h]
   \includegraphics[width = 0.4\textwidth]{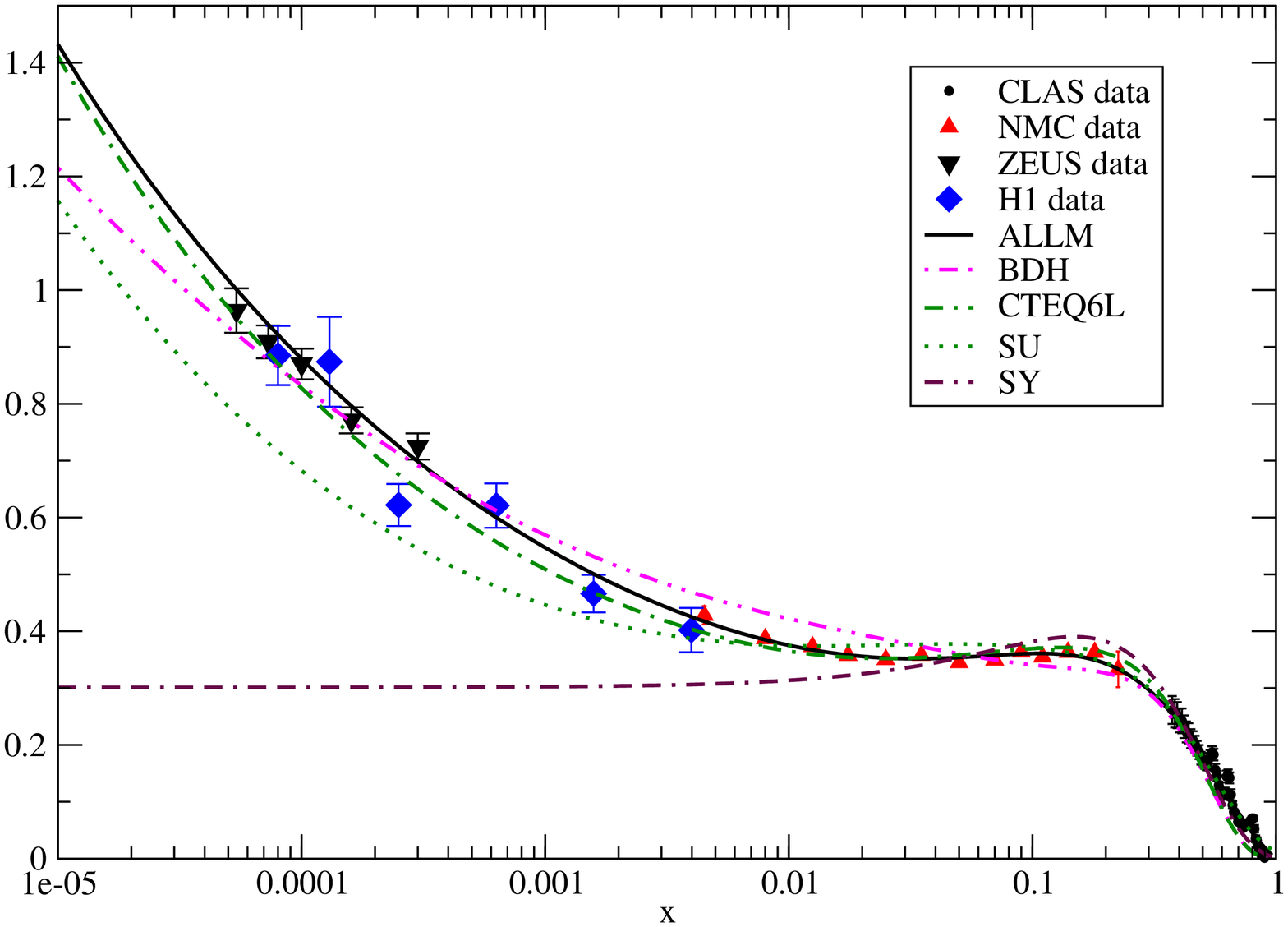}
    \caption{The proton structure function $F_2(x,Q^2)$ as a function of $x$ at $Q^2 = 2.5 \, \rm{GeV}^2$, shown with a logarithmic $x$-axis, to
make visible the small-$x$ behaviour of different parametrizations. Here also HERA data are included.
 }
  \label{fig:F2_lowx}
 \end{figure}

In Fig.\ref{fig:F2} we show the proton structure function $F_2(x,Q^2)$
obtained from the various fits at $Q^2= 0.225, 1.25, 2.5, 4.5  \, \rm{GeV}^2$ 
as a function of Bjorken-$x$ in Fig. \ref{fig:F2}. In
Fig. \ref{fig:F2_lowx}, we show the structure function 
$F_2(x,Q^2)$ at $Q^2= 2.5 \rm{GeV}^2$,
but this time with a logarithmic abscissa to emphasize the low-$x$
behaviour of different parametrizations. 
Also shown are the HERA data at low-$x$. 
Experimental data on the figures are taken from the compilation
\cite{Gehrmann:1999xn} and from \cite{Osipenko:2003ua,Osipenko:2003bu}.

Here we see that the Suri-Yennie fit corresponds to a unit-intercept 
Pomeron and does not describe the small-$x$ rise of the proton structure function.

A surprising lesson is, that the old Suri-Yennie \cite{Suri:1971yx} fit,
still gives a reasonable description of $F_2$ except of very small $x$.

For explicit account of resonances it would be recommended to use the
Fiore et al. \cite{Fiore:2002re}, but care has to be taken
to stay within the resonance region, as the quality of the fit beyond
this region quickly deteriorates. 

The overall best description appears to be given by the ALLM 
\cite{Abramowicz:1991xz,Abramowicz:1997ms} fit.

\subsection{Monte Carlo generator}

In contrast to our previous studies \cite{daSilveira:2014jla}, all calculations
performed within the $k_T$-factorization approach were performed with a Monte
Carlo event generator, where the formulae presented
above (see also \cite{daSilveira:2014jla}) were implemented. 
This Monte Carlo program is used to generate events (four-momenta of
leptons and outgoing protons/excited systems) which are then transformed to distributions
with the help of the standard software Root \cite{Root}.
The typical number of events generated in our studies is a few millions.
A more detailed description of the event generator will be presented elsewhere 
\cite{MC_generator}.

\section{Results}
\label{sec:results}

\begin{figure}
\begin{minipage}{0.45\textwidth}
 \centerline{\includegraphics[width=1.0\textwidth]{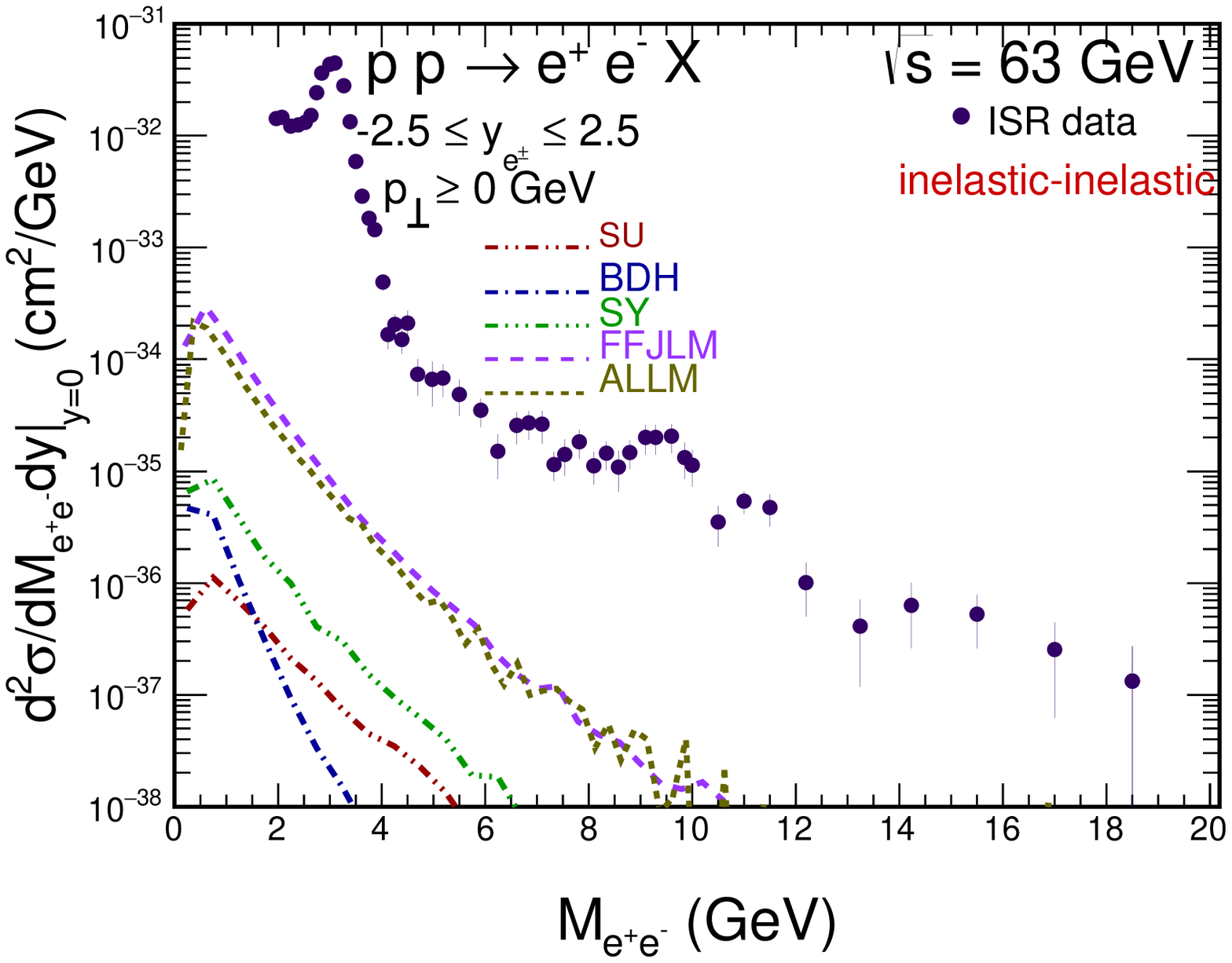}}
\end{minipage}
\hspace{0.2cm}
\begin{minipage}{0.45\textwidth}
 \centerline{\includegraphics[width=1.0\textwidth]{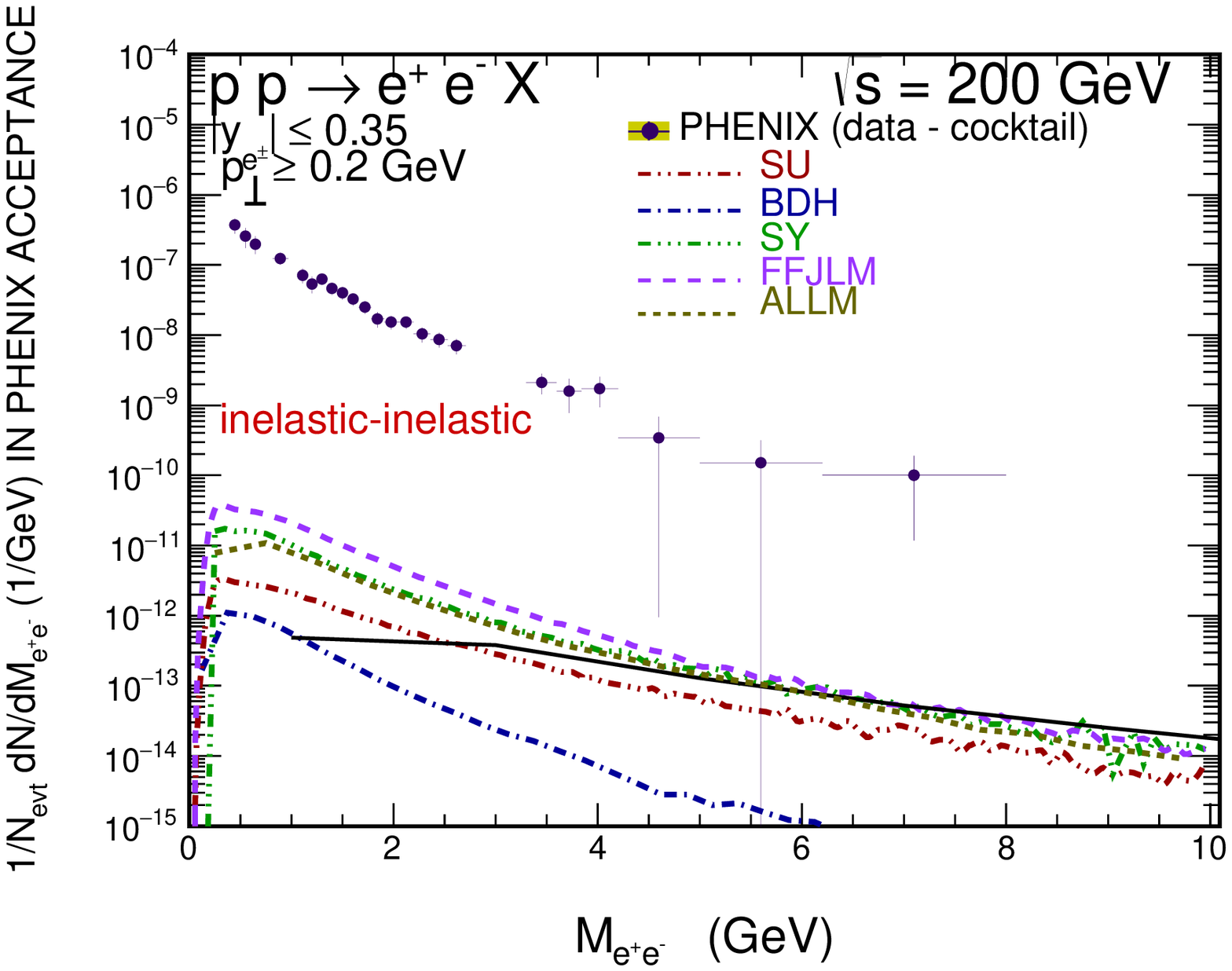}}
\end{minipage}
\hspace{0.2cm}
\begin{minipage}{0.45\textwidth}
 \centerline{\includegraphics[width=1.0\textwidth]{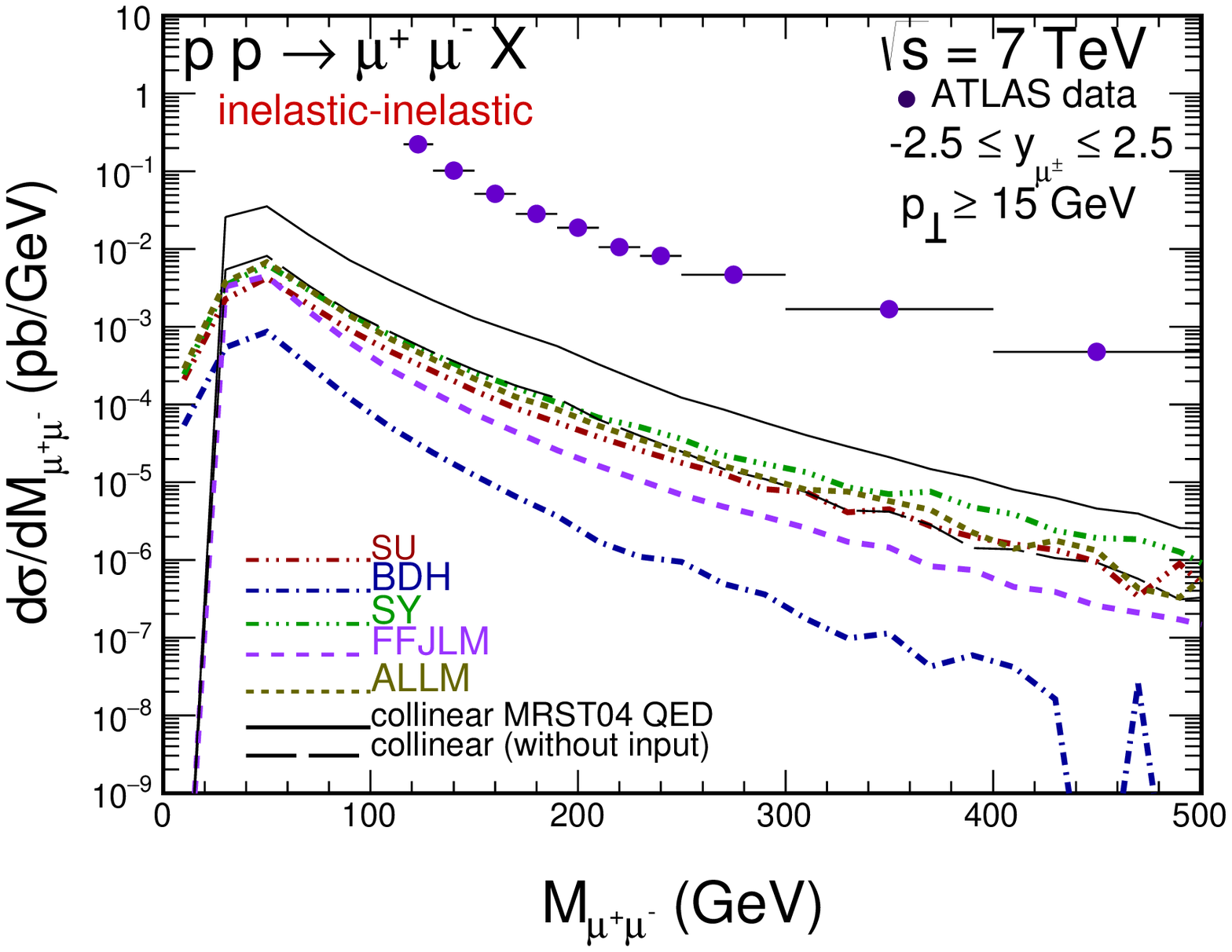}}
\end{minipage}
\hspace{0.2cm}
\begin{minipage}{0.45\textwidth}
 \centerline{\includegraphics[width=1.0\textwidth]{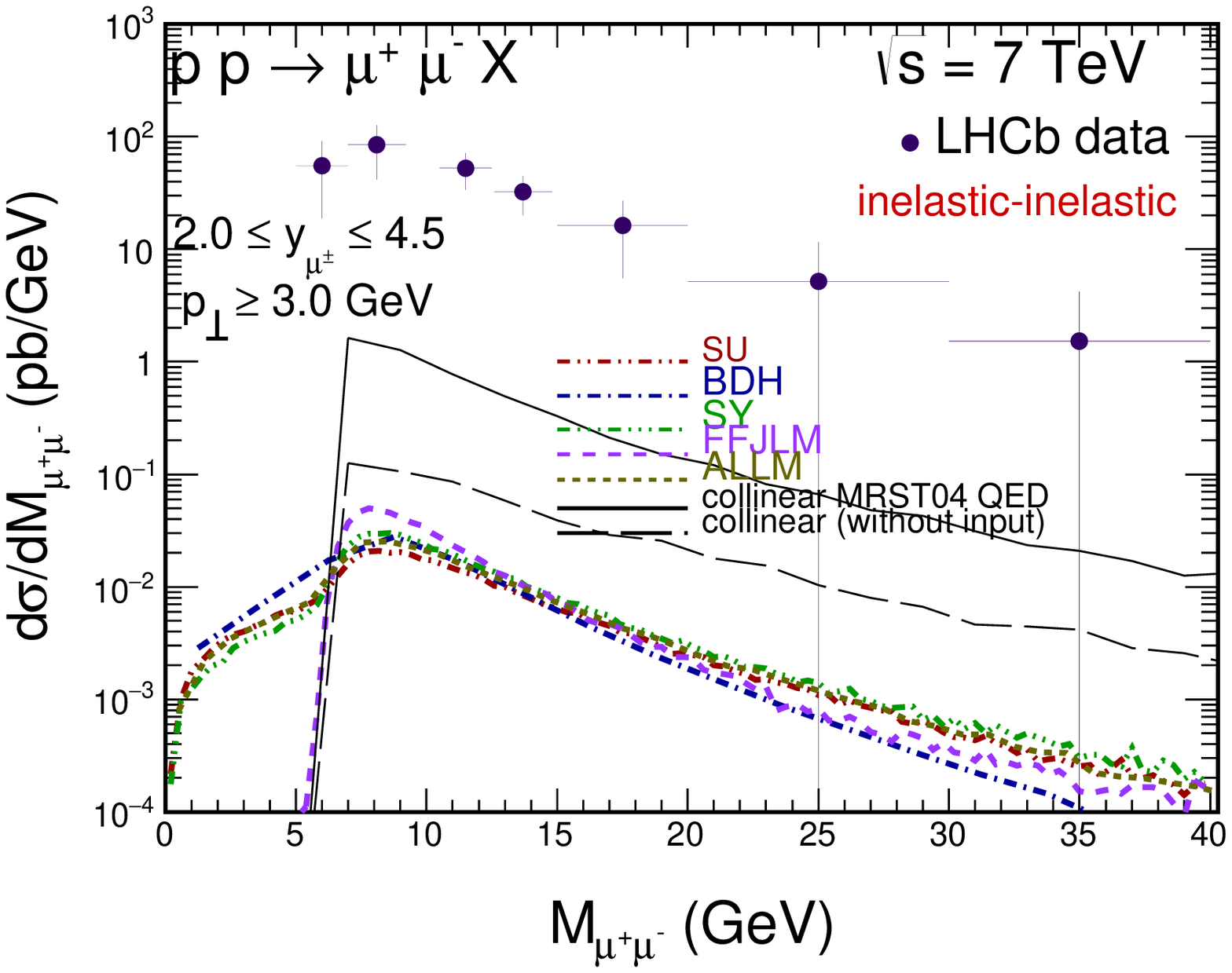}}
\end{minipage}
   \caption{The inelastic-inelastic contribution to dilepton invariant mass 
distributions for the ISR (upper-left), PHENIX (upper-right), 
ATLAS (lower-left) and LHCb (lower-right) experiments 
for different structure functions.}
 \label{fig:dsig_dMll_ine_ine}
\end{figure}

Most of the experiments for the dilepton production concentrate on
determination of dilepton invariant mass distributions. 
In Fig.\ref{fig:dsig_dMll_ine_ine} we show invariant mass distributions
of dilepton pairs produced in the photon-photon inelastic-inelastic 
mechanism for kinematical conditions relevant for different experiments.
We show results obtained with different parametrizations of the structure
functions known from the literature. Surprisingly the different
structure functions give quite different results.
For completeness in some cases (when possible) we also show 
the result obtained in the collinear approach with the MRST2004(QED) 
photon distribution with (solid black line)
and similar one when ignoring the initial input (long-dashed black line). 
The result obtained within the collinear approach with the 
MRST2004(QED) distribution is much above the results obtained within 
the $k_T$-factorization approach.
In our opinion this is mainly related to the large input photon distribution
at the initial scale $Q_0^2$ = 2 GeV$^2$ (see Eq.(\ref{photon_initial})) discussed 
in the context of Fig.\ref{fig:collinear_photon_pdf}.
If the input is discarded (long-dashed black line) the collinear
result is similar to the results obtained within the $k_T$-factorization.
The inelastic-inelastic contribution gives only a small fraction
of the measured cross section for most experimental conditions 
(ATLAS,LHCb,PHENIX). For the ISR experiment it is relatively larger.

\begin{figure}
\begin{minipage}{0.45\textwidth}
 \centerline{\includegraphics[width=1.0\textwidth]{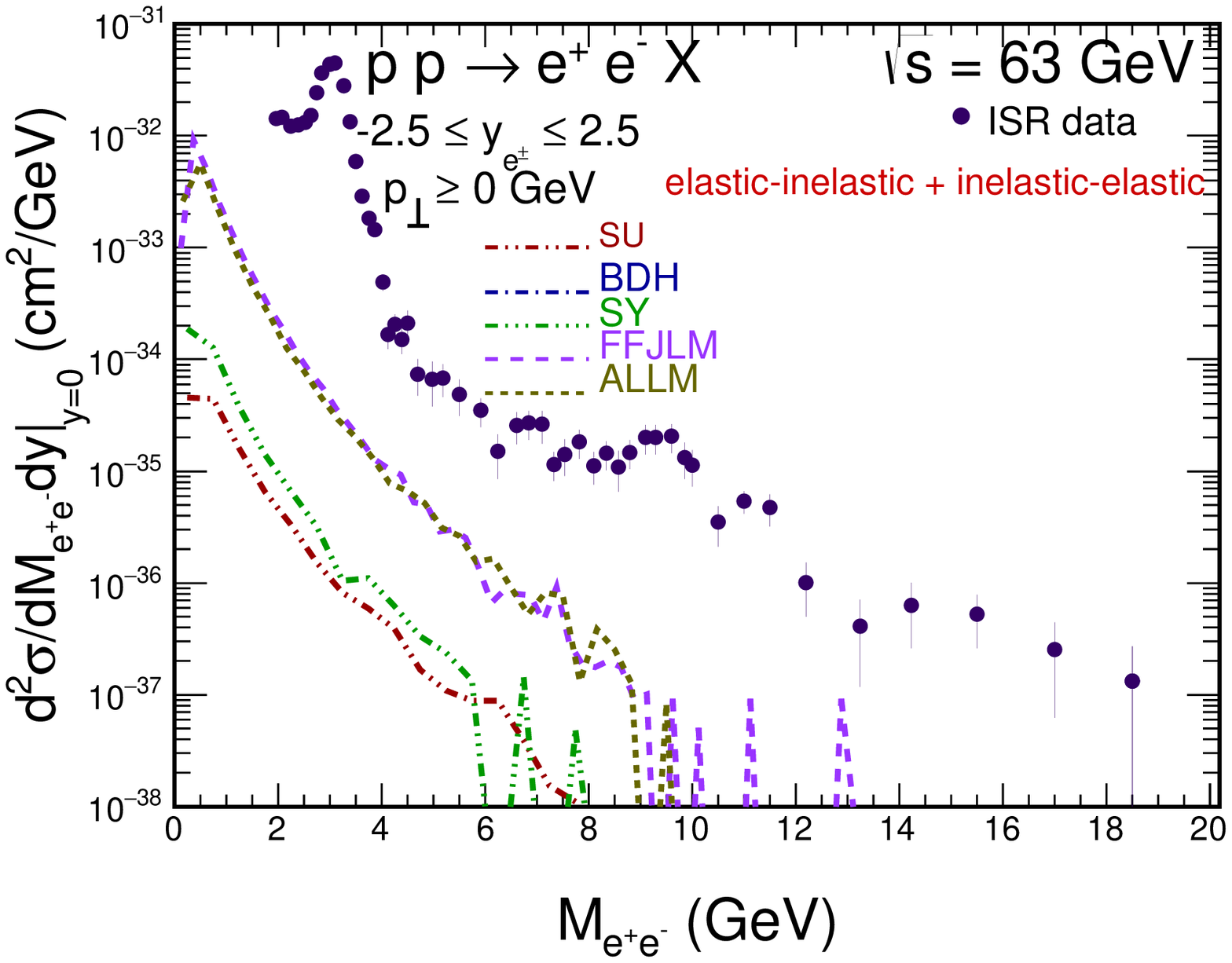}}
\end{minipage}
\hspace{0.2cm}
\begin{minipage}{0.45\textwidth}
 \centerline{\includegraphics[width=1.0\textwidth]{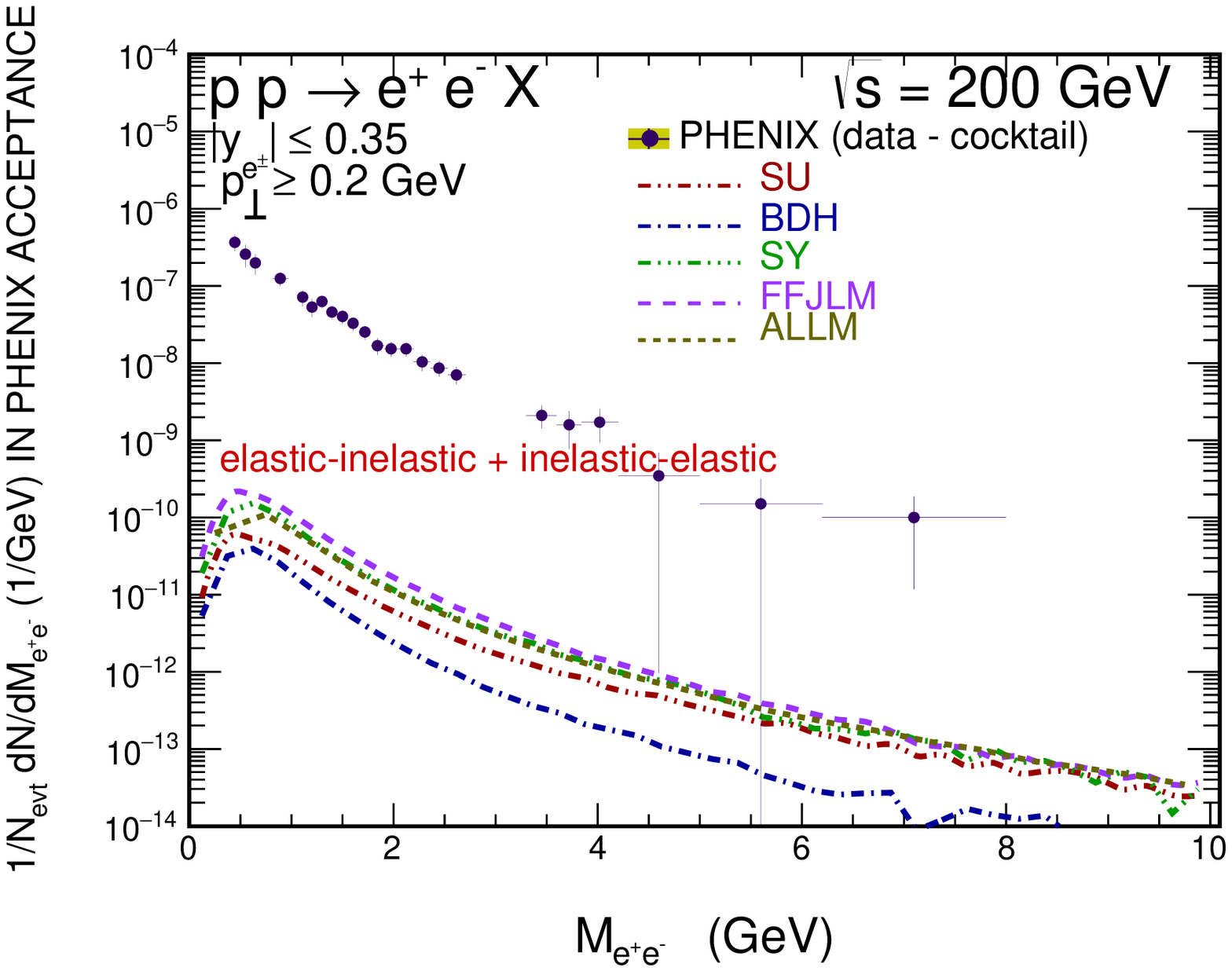}}
\end{minipage}
\hspace{0.2cm}
\begin{minipage}{0.45\textwidth}
 \centerline{\includegraphics[width=1.0\textwidth]{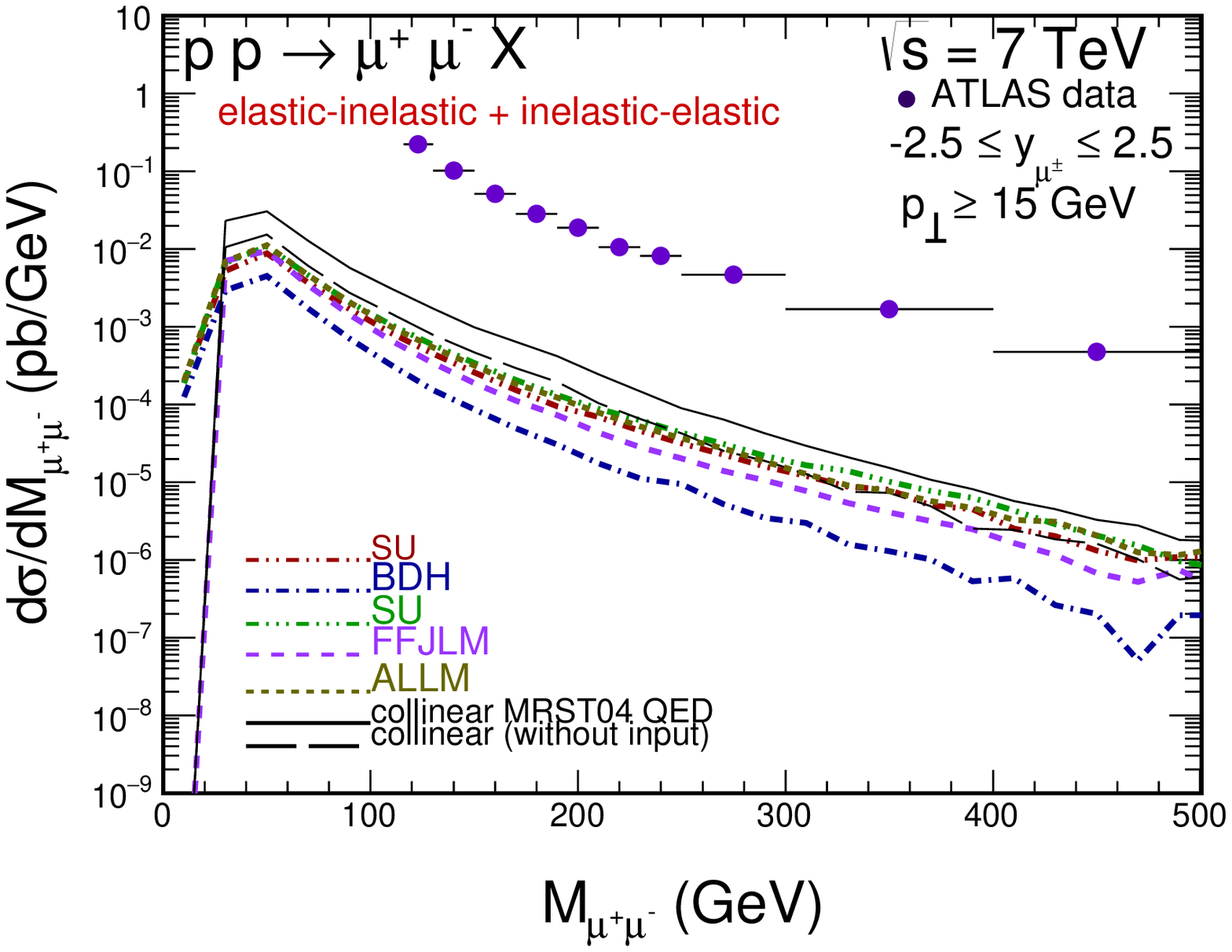}}
\end{minipage}
\hspace{0.2cm}
\begin{minipage}{0.45\textwidth}
 \centerline{\includegraphics[width=1.0\textwidth]{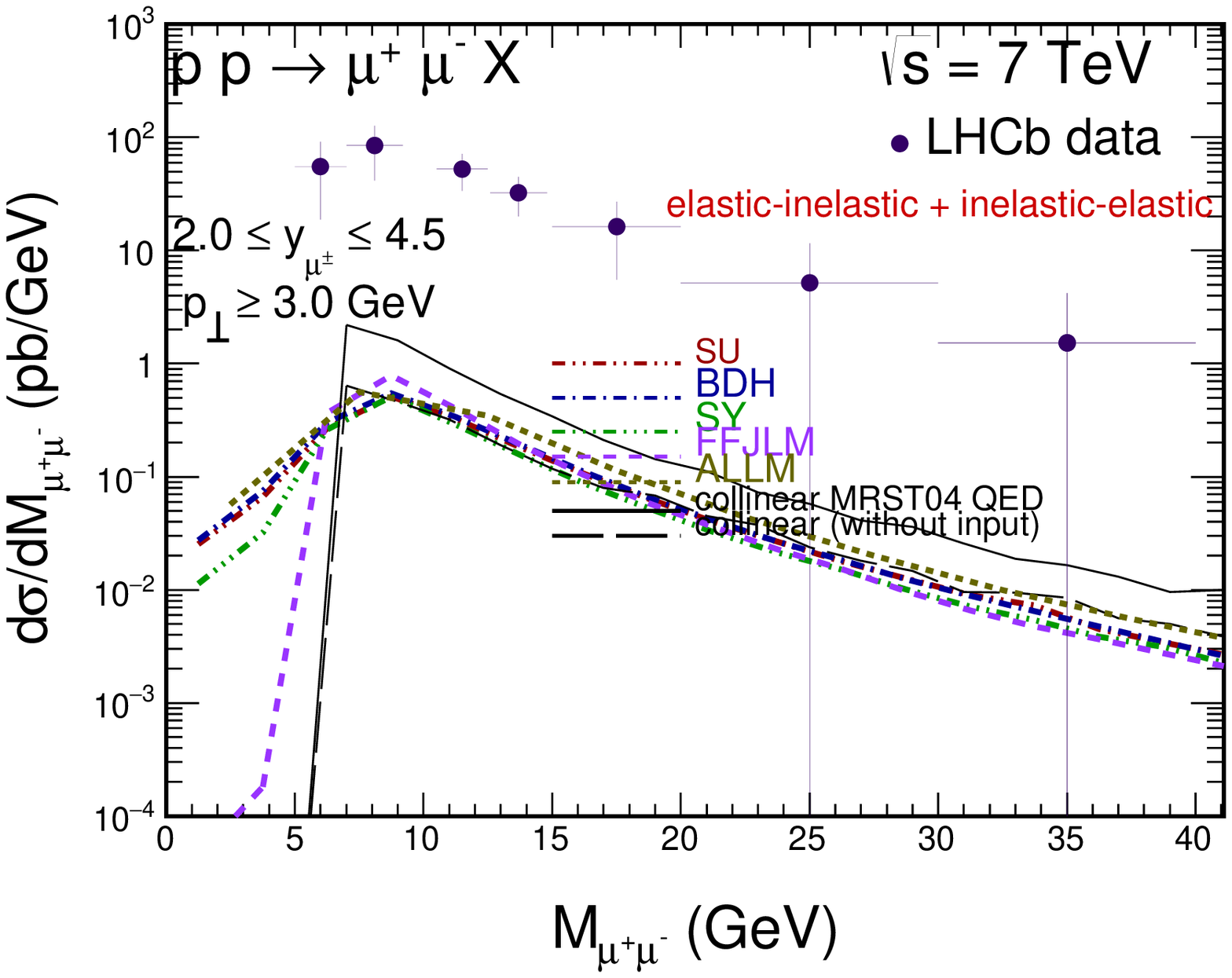}}
\end{minipage}
   \caption{The (elastic-inelastic)+(inelastic-elastic) contribution 
to dilepton invariant mass distributions for the ISR (upper-left), 
PHENIX (upper-right), ATLAS (lower-left) and LHCb (lower-right) experiments 
for different structure functions.}
 \label{fig:dsig_dMll_ela_ine}
\end{figure}

In Fig.\ref{fig:dsig_dMll_ela_ine} we show dilepton invariant mass 
distributions for elastic-inelastic and inelastic-elastic (added
together) contributions.
As for inelastic-inelastic contribution the results strongly depend
on the parametrization of the structure functions used.
The spread of results for different $F_2$ from the literature
is now, however, significantly smaller than in the case of 
inelastic-inelastic contributions where the structure functions enter
twice (into both photon flux factors).
As for the double inelastic case we also show a result for
the collinear approach.
The mixed components give similar contribution to the dilepton
invariant mass distributions as the inelastic-inelastic one.

\begin{figure}[!h]
\begin{minipage}{0.45\textwidth}
 \centerline{\includegraphics[width=1.0\textwidth]{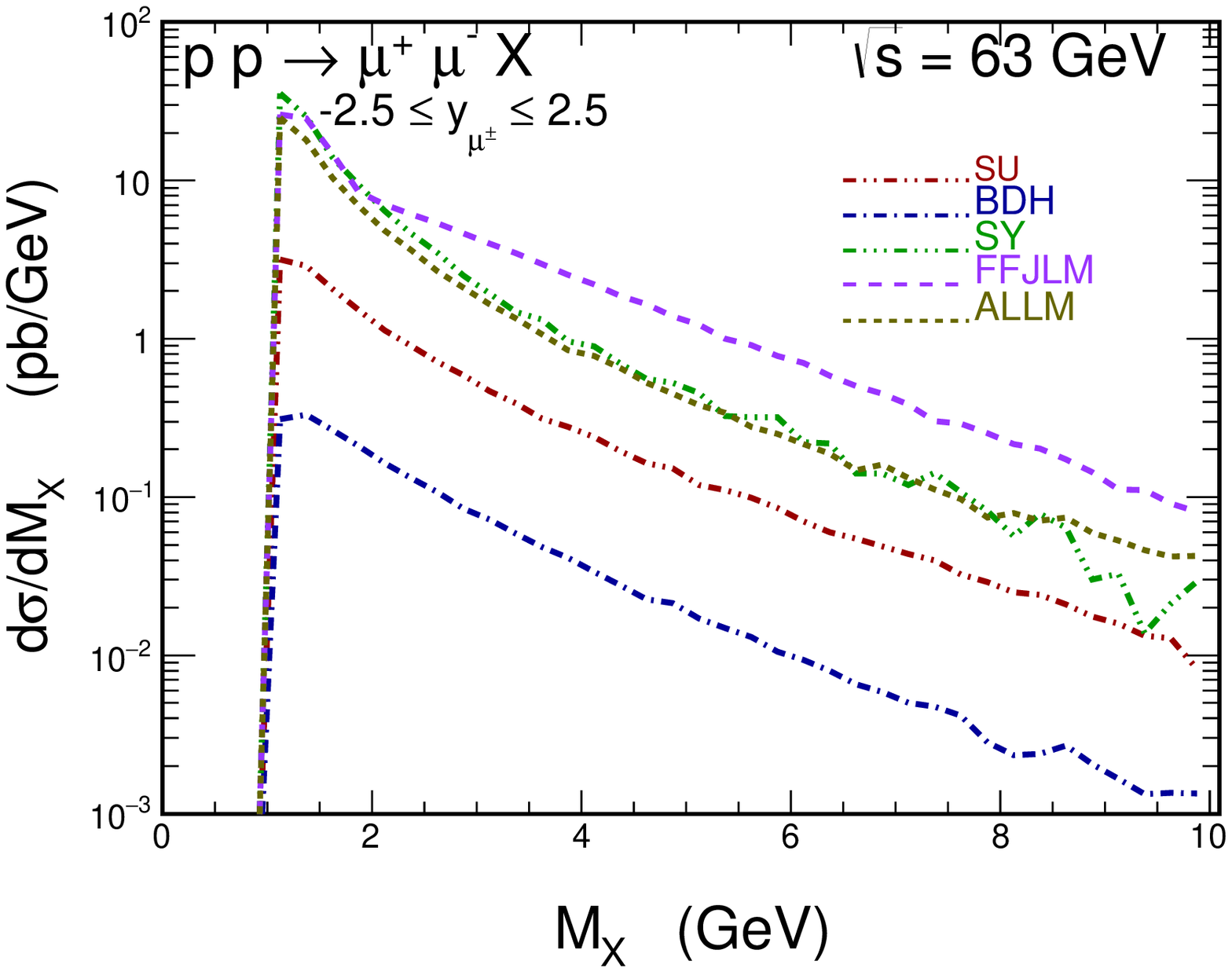}}
\end{minipage}
\hspace{0.2cm}
\begin{minipage}{0.45\textwidth}
 \centerline{\includegraphics[width=1.0\textwidth]{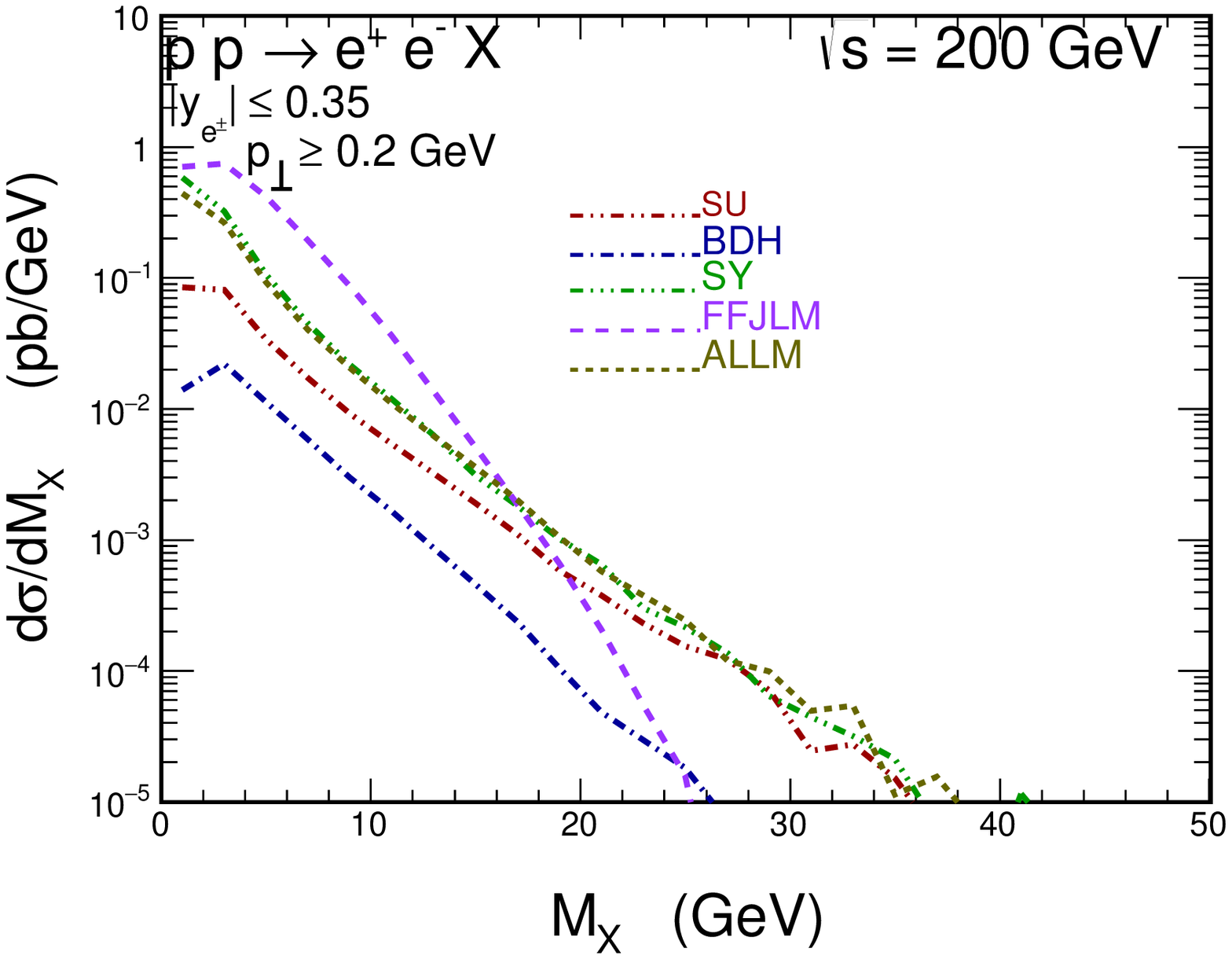}}
\end{minipage}
\hspace{0.2cm}
\begin{minipage}{0.45\textwidth}
 \centerline{\includegraphics[width=1.0\textwidth]{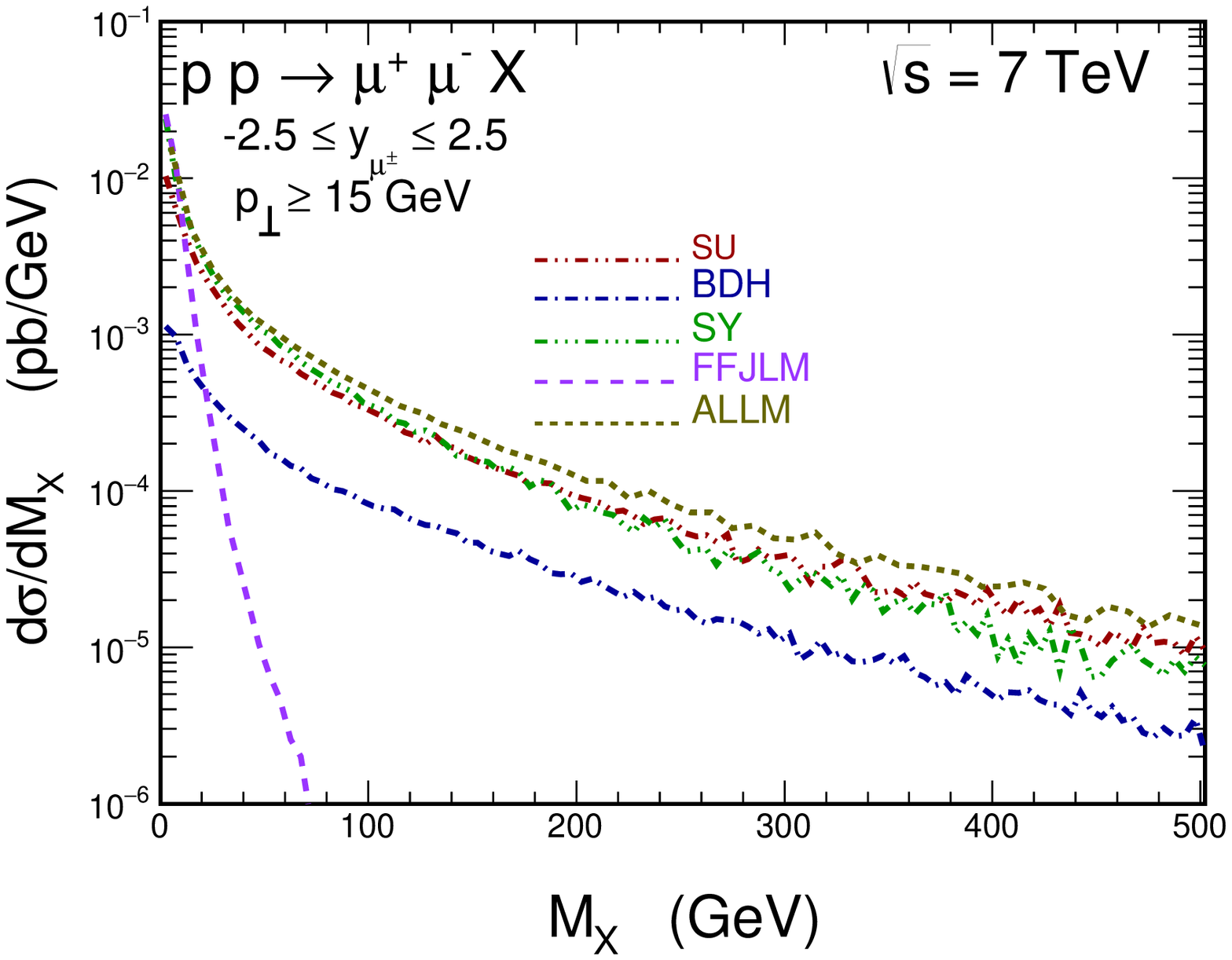}}
\end{minipage}
\hspace{0.2cm}
\begin{minipage}{0.45\textwidth}
 \centerline{\includegraphics[width=1.0\textwidth]{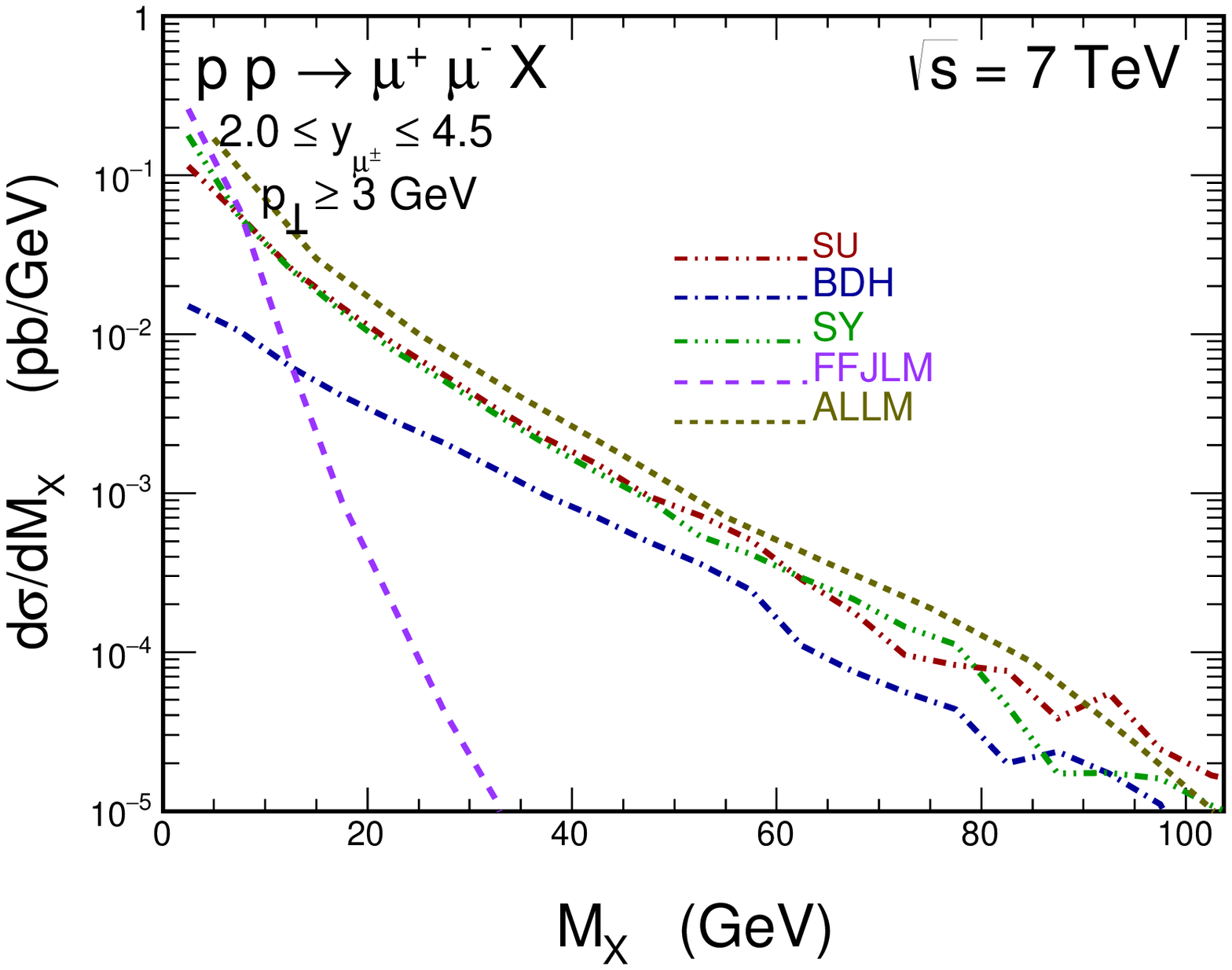}}
\end{minipage}
   \caption{Missing mass distributions for ineleastic-inelastic 
   photon-photon contributions for different experiments 
   (ISR, PHENIX, ATLAS, LHCb)
and different parametrizations of the structure functions
as explained inside the figures.
}
 \label{fig:dsig_dMX_ine_ine}
\end{figure}

It is very interesting to understand which regions of
($Q_1^2, M_X$) and ($Q_2^2, M_Y$) space contribute in the measured spectra.
We start our review from distributions in $M_X$ (or $M_Y$).
The corresponding results for the inelastic-inelastic component
are shown in Fig.\ref{fig:dsig_dMX_ine_ine}. Again we show results for 
the ISR (left top panel), PHENIX (right top panel)
ATLAS (left bottom panel) and LHCb (right bottom panel)  experiment.
The dominant contributions come from the region of very small
missing masses $M_X$ (or $M_Y$). This is not necessarily the region 
where standard evolution equation applies for the description of 
$F_2$ structure function. In general, the Fiore et al. \cite{Fiore:2002re} and 
Suri-Yennie \cite{Suri:1971yx} parametrization give much bigger cross section
in the region of small missing masses.
In this plot the resolution in missing mass is rather coarse 
($\Delta M_X$ = 2.5 GeV).
If the resolution of the distribution (binning) was improved one 
could observe even peaks corresponding to nucleon resonances excited 
by virtual photons for the Fiore et al. parametrization. 
As seen in Fig.\ref{fig:F2} the Suri-Yennie parametrization extremely 
well averages the structures 
in more detailed Fiore et al. parametrization.
Clearly the Fiore et al. parametrization is not adequate for
large $M_X$ ($M_Y$) masses.
All this demonstrates how important is using a ``proper'' structure
function.

\begin{figure}[!h]
\begin{minipage}{0.45\textwidth}
 \centerline{\includegraphics[width=1.0\textwidth]{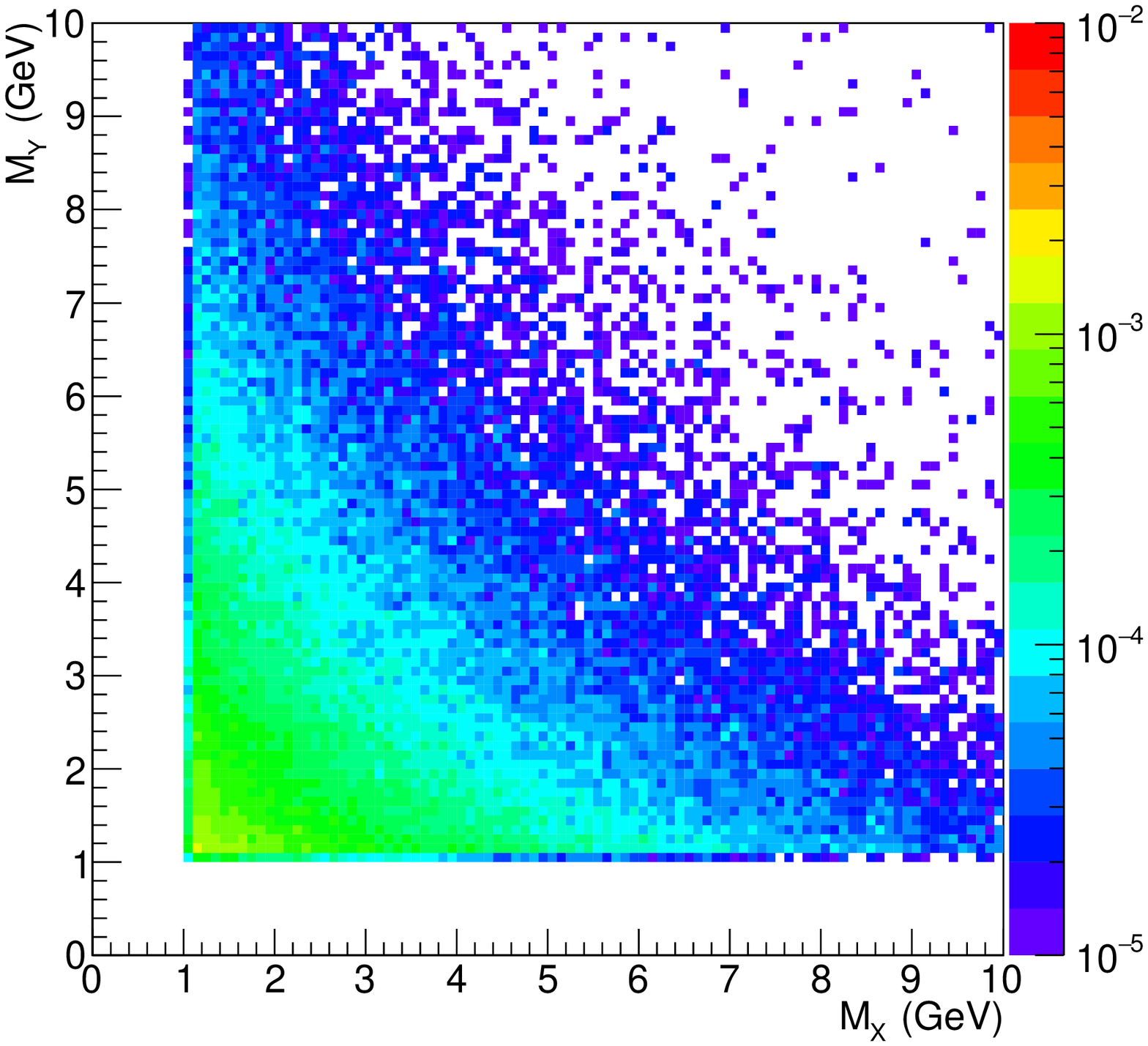}}
\end{minipage}
\hspace{0.2cm}
\begin{minipage}{0.45\textwidth}
 \centerline{\includegraphics[width=1.0\textwidth]{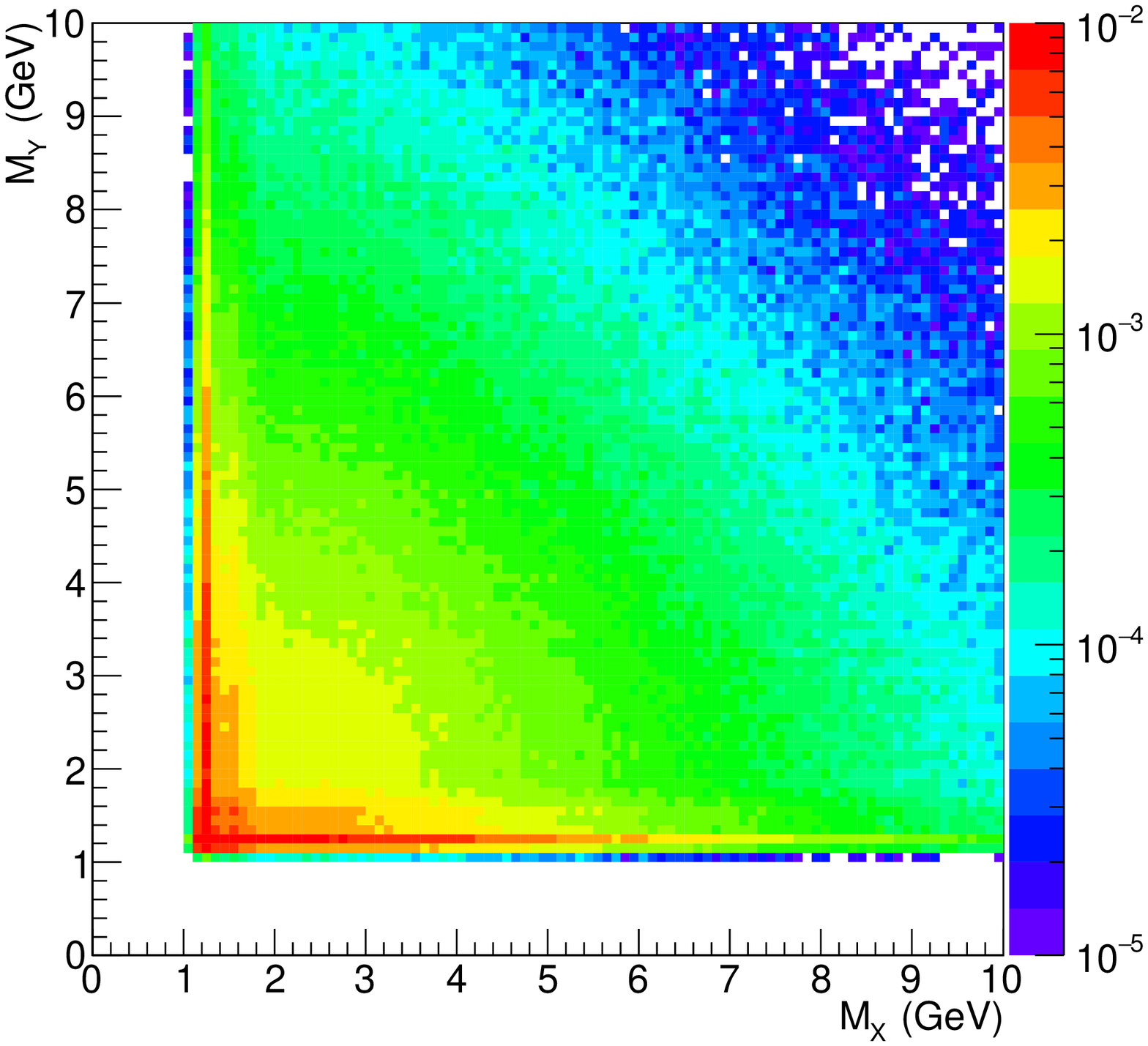}}
\end{minipage}
\hspace{0.2cm}
\begin{minipage}{0.45\textwidth}
 \centerline{\includegraphics[width=1.0\textwidth]{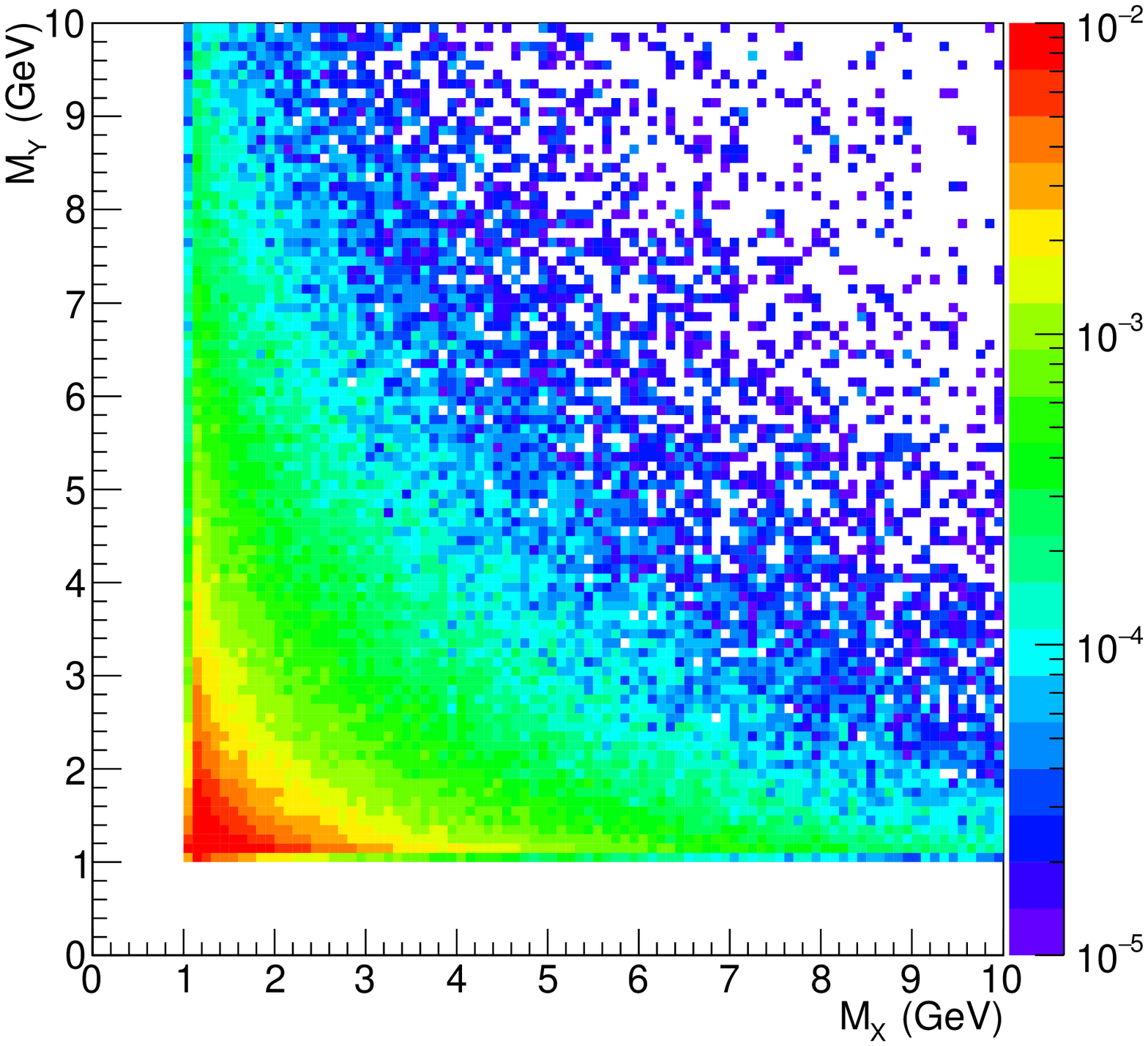}}
\end{minipage}
\hspace{0.2cm}
\begin{minipage}{0.45\textwidth}
 \centerline{\includegraphics[width=1.0\textwidth]{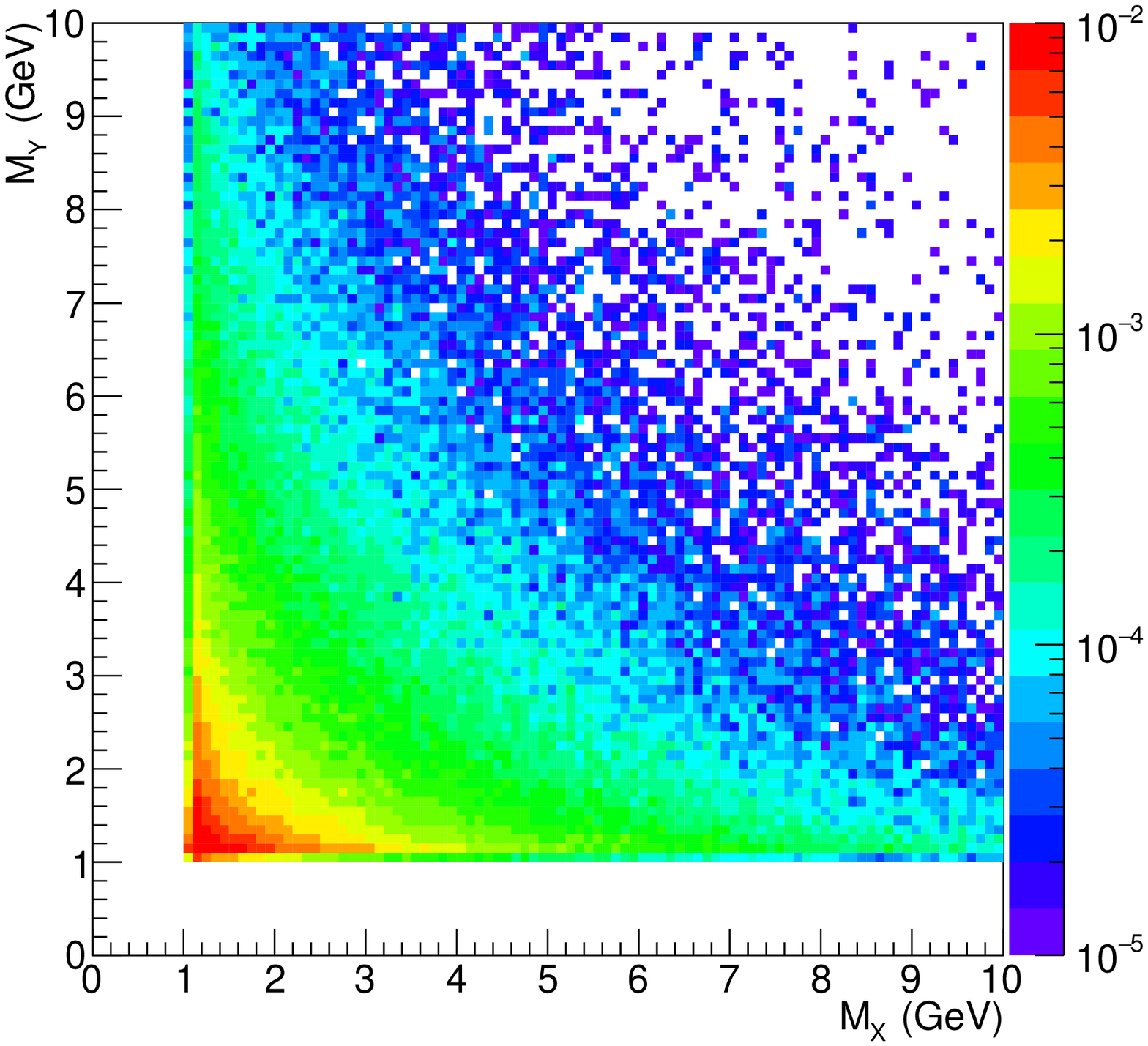}}
\end{minipage}
   \caption{Distributions for $M_X \times M_Y$ for different 
structure functions:
Szczurek-Uleshchenko (upper-left), 
Fiore et al. (upper-right),
Suri-Yennie (lower-right) and 
ALLM (lower-left)
for $\sqrt{s}$ = 200 GeV and $p_{T1}, p_{T2} >$ 0.2 GeV.
}
 \label{fig:dsig_dMXdMY}
\end{figure}

In Fig.\ref{fig:dsig_dMXdMY} we show some examples of two-dimensional 
distributions $(M_X, M_Y)$ for different parametrizations of 
the structure functions as an example for the PHENIX kinematics.
Here we focus on small values of $M_X$ and $M_Y$ to resolve apparent
differences.
Clearly the different parametrizations give very different results.
In the case of Fiore et al. parametrization one can observe now (with
better resolution) resonance lines for $M_X$ or $M_Y$ slightly bigger
than 1 GeV.

\begin{figure}[!h]
\begin{minipage}{0.45\textwidth}
 \centerline{\includegraphics[width=1.0\textwidth]{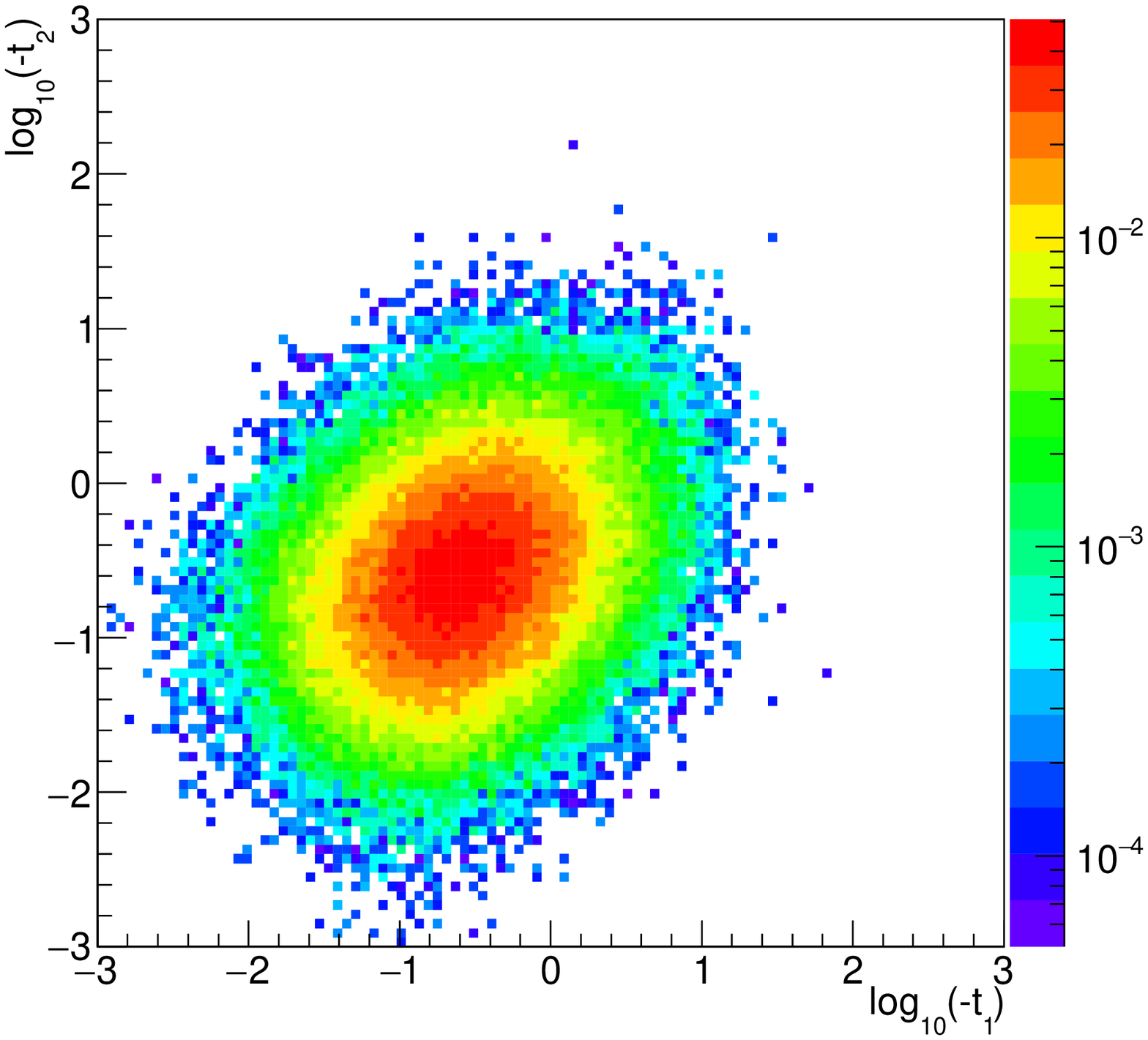}}
\end{minipage}
\hspace{0.2cm}
\begin{minipage}{0.45\textwidth}
 \centerline{\includegraphics[width=1.0\textwidth]{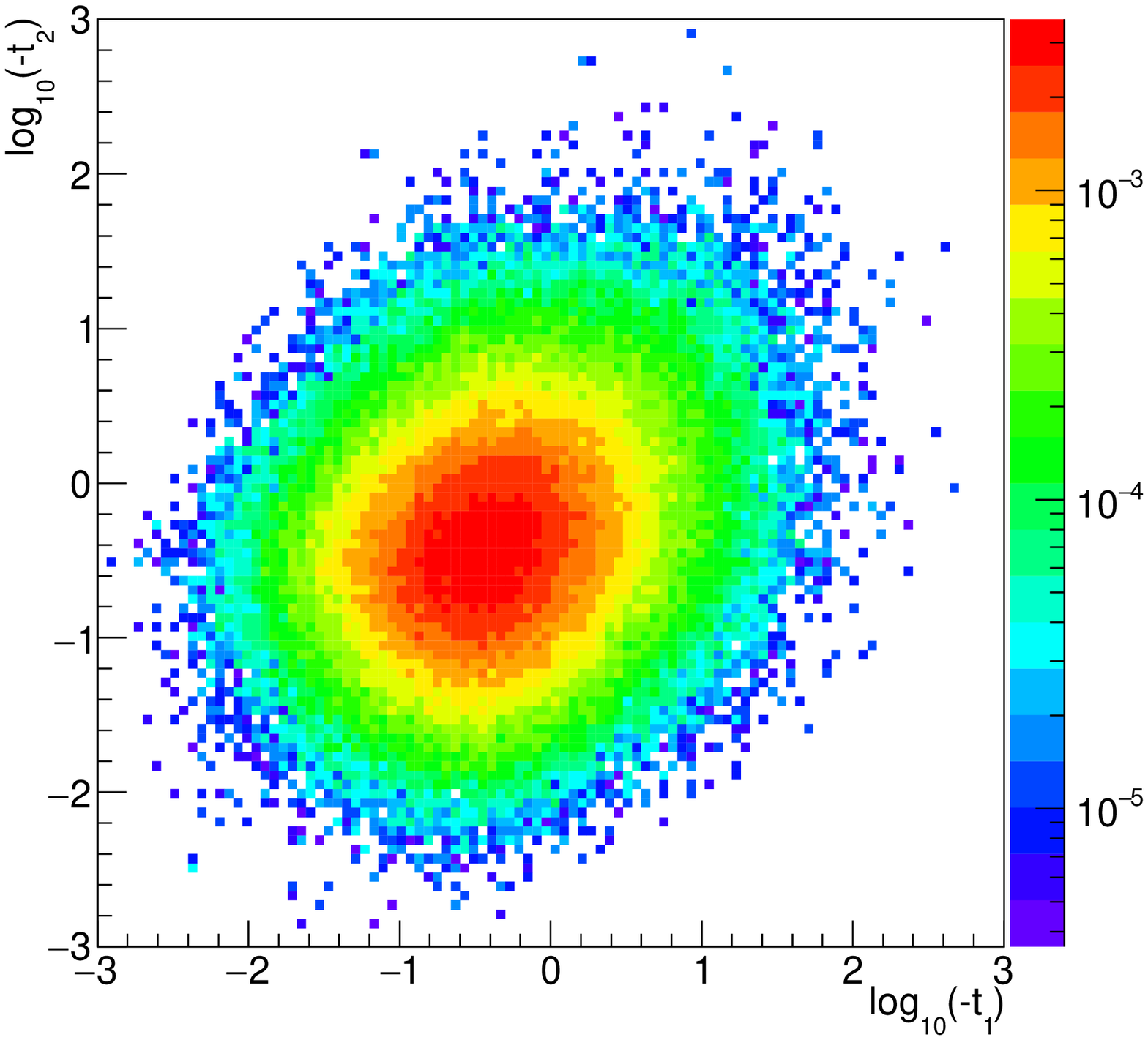}}
\end{minipage}
\hspace{0.2cm}
\begin{minipage}{0.45\textwidth}
 \centerline{\includegraphics[width=1.0\textwidth]{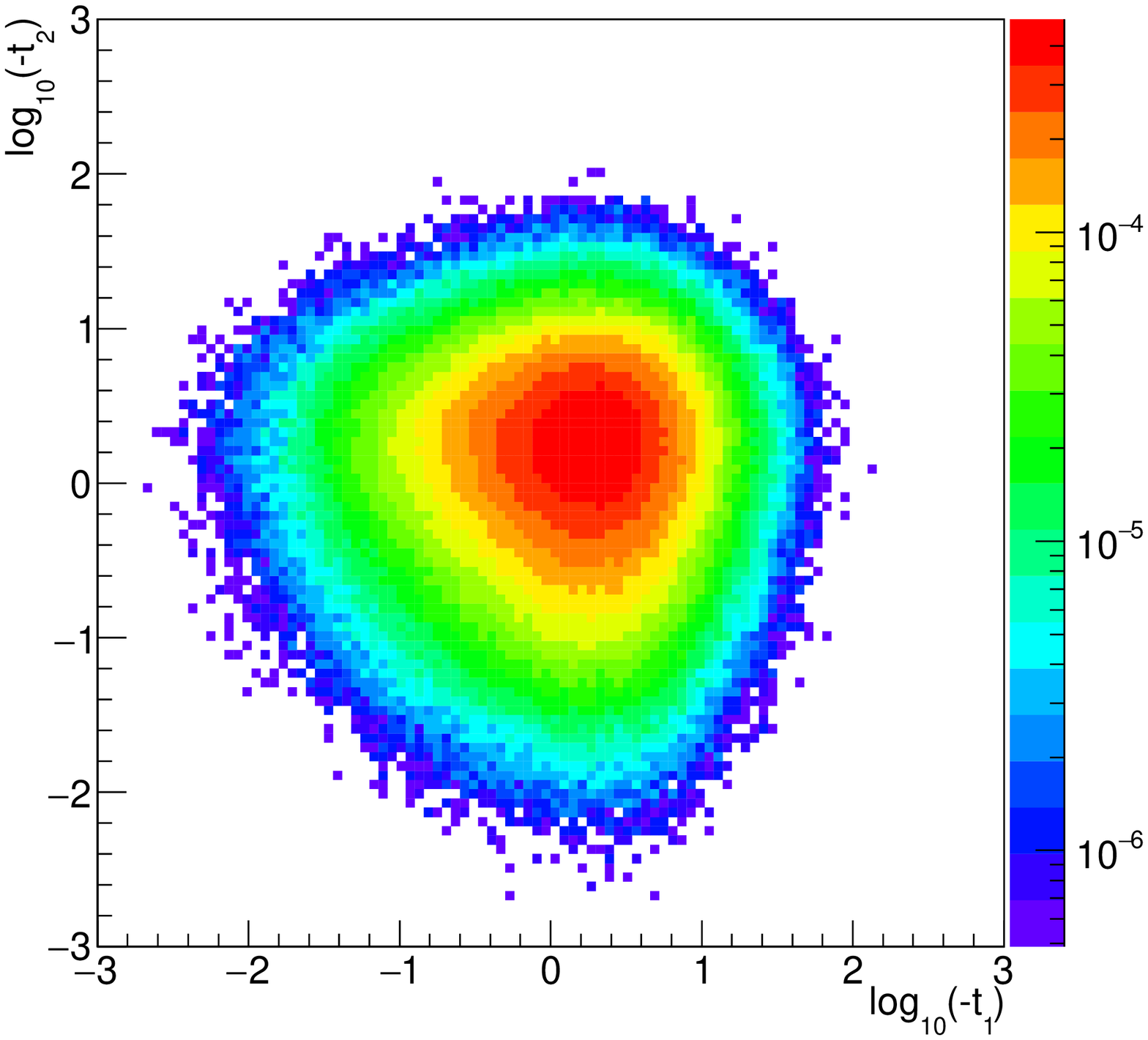}}
\end{minipage}
\hspace{0.2cm}
\begin{minipage}{0.45\textwidth}
 \centerline{\includegraphics[width=1.0\textwidth]{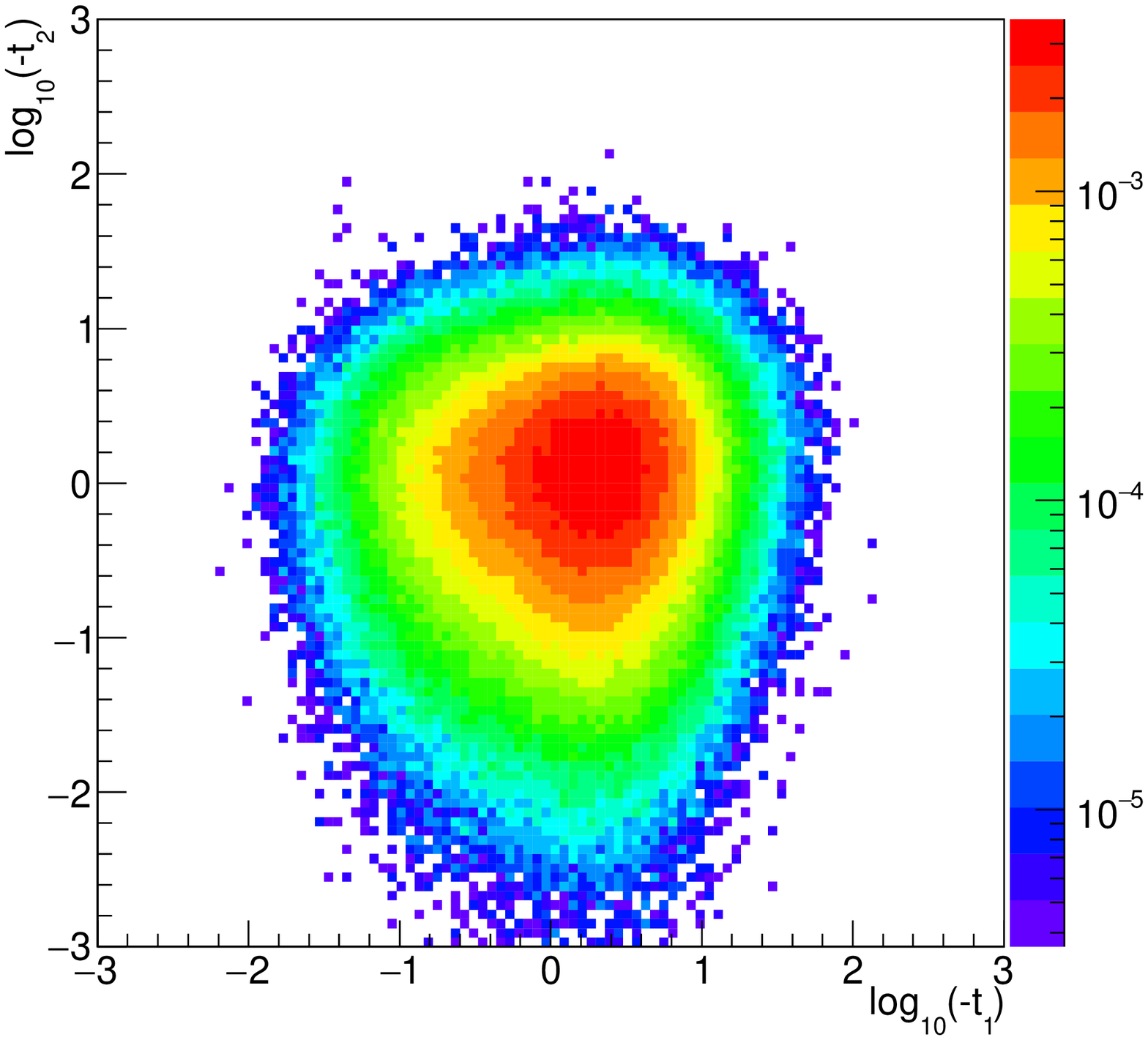}}
\end{minipage}
   \caption{Distributions for $Q_1^2 \times Q_2^2$ for different 
experiments:
ISR (upper-left), 
PHENIX (upper-right),
ATLAS (lower-right) and 
LHCb (lower-left)
for ALLM structure function.
}
\label{fig:dsig_dQ12dQ22}
\end{figure}

In Fig.\ref{fig:dsig_dQ12dQ22} we show two-dimensional distributions
$(Q_1^2,Q_2^2)$ for four different experimental conditions
specified in the figure caption.
In most of the cases rather large photon virtualities contribute.
This is especially true for the ATLAS experiment with large cuts
on lepton transverse momenta \cite{ATLAS_2013_high-mass}. In the case of 
the old ISR experiment \cite{ISR} or more recent PHENIX experiment 
\cite{PHENIX} the situation is very different and clearly contributions 
from $F_2$ nonperturbative regions come into game and should 
be carefully analyzed.
For the case of LHCb, in contrast to other cases, the distribution in
$Q_1^2 \times Q_2^2$ is not symmetric along the diagonal which is
related to asymmetric forward coverage of the LHCb experiment.

Summarizing this part, we have shown that with typical experimental
cuts the contribution of photon-photon fusion is much smaller
than dilepton experimental data and constitutes typically less than
1 \% of the measured cross sections.

\begin{figure}
\begin{center}
\includegraphics[width=8cm]{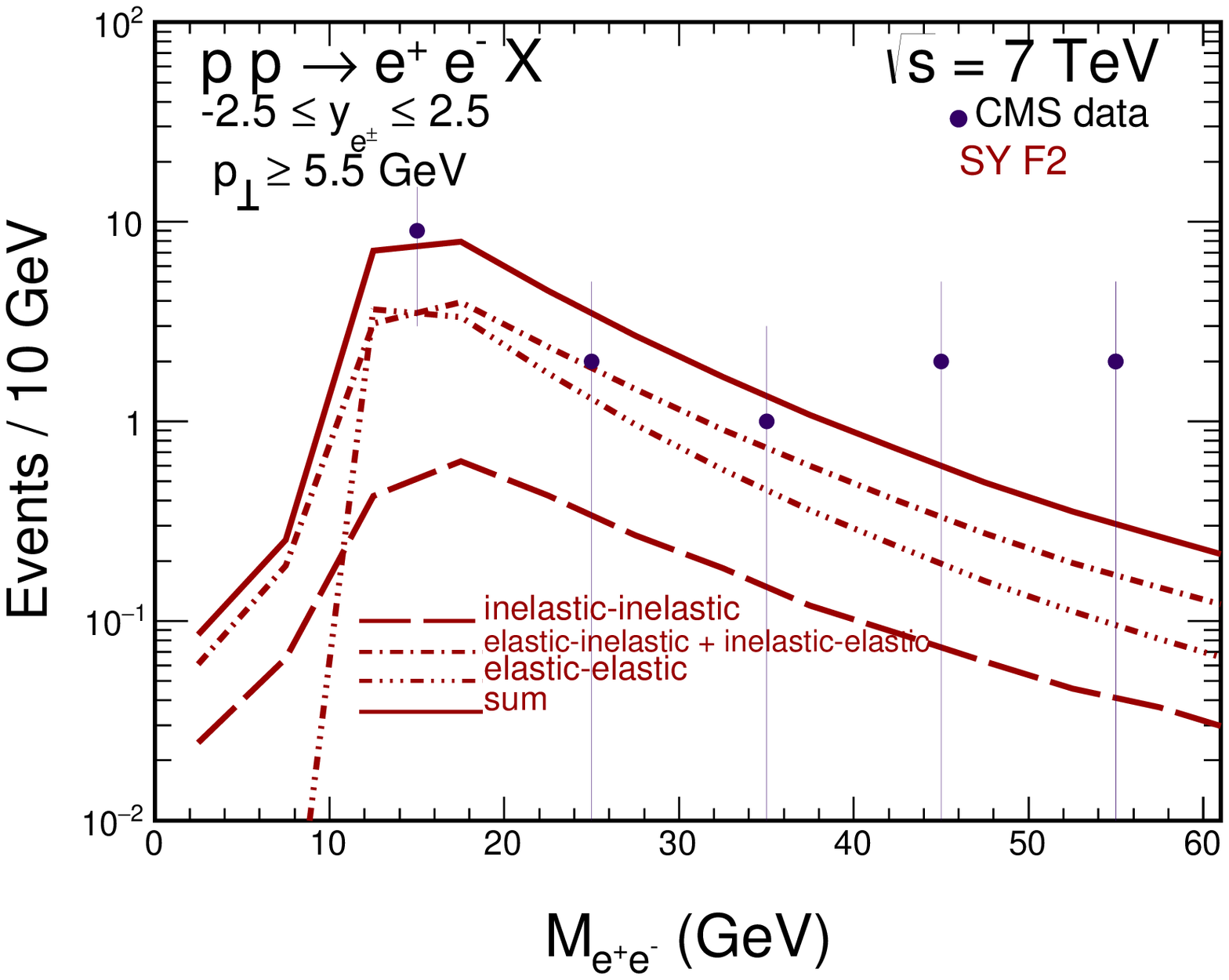}
\includegraphics[width=8cm]{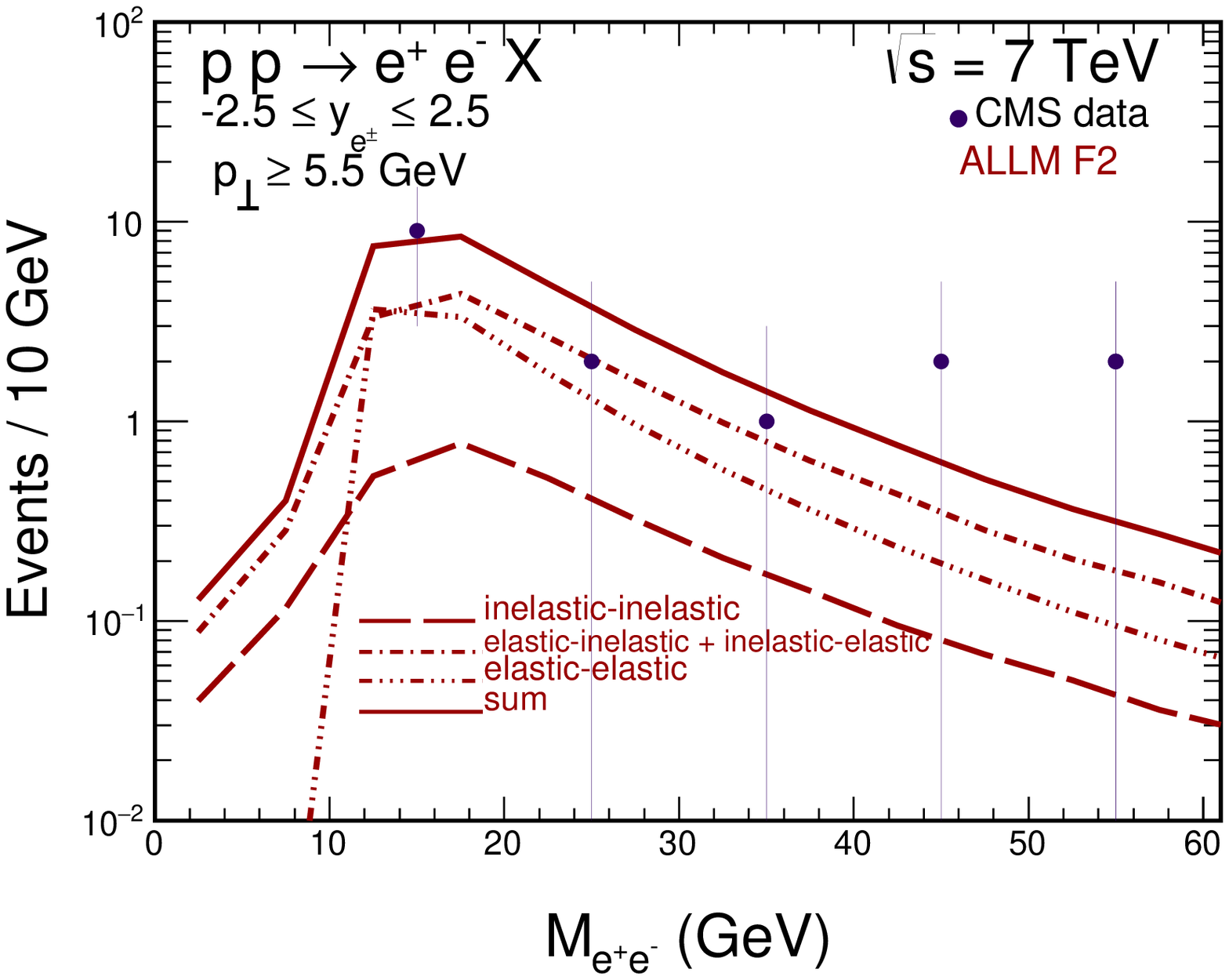}
\end{center}
\caption{
Number of events per invariant mass interval for the CMS experimental
cuts for SY (left) and ALLM (right) structure functions.
The experimental data points are from Ref.\cite{Chatrchyan:2012tv}.
}
\label{fig:CMS_dN_dMll}
\end{figure}

\begin{figure}
\begin{center}
\includegraphics[width=8cm]{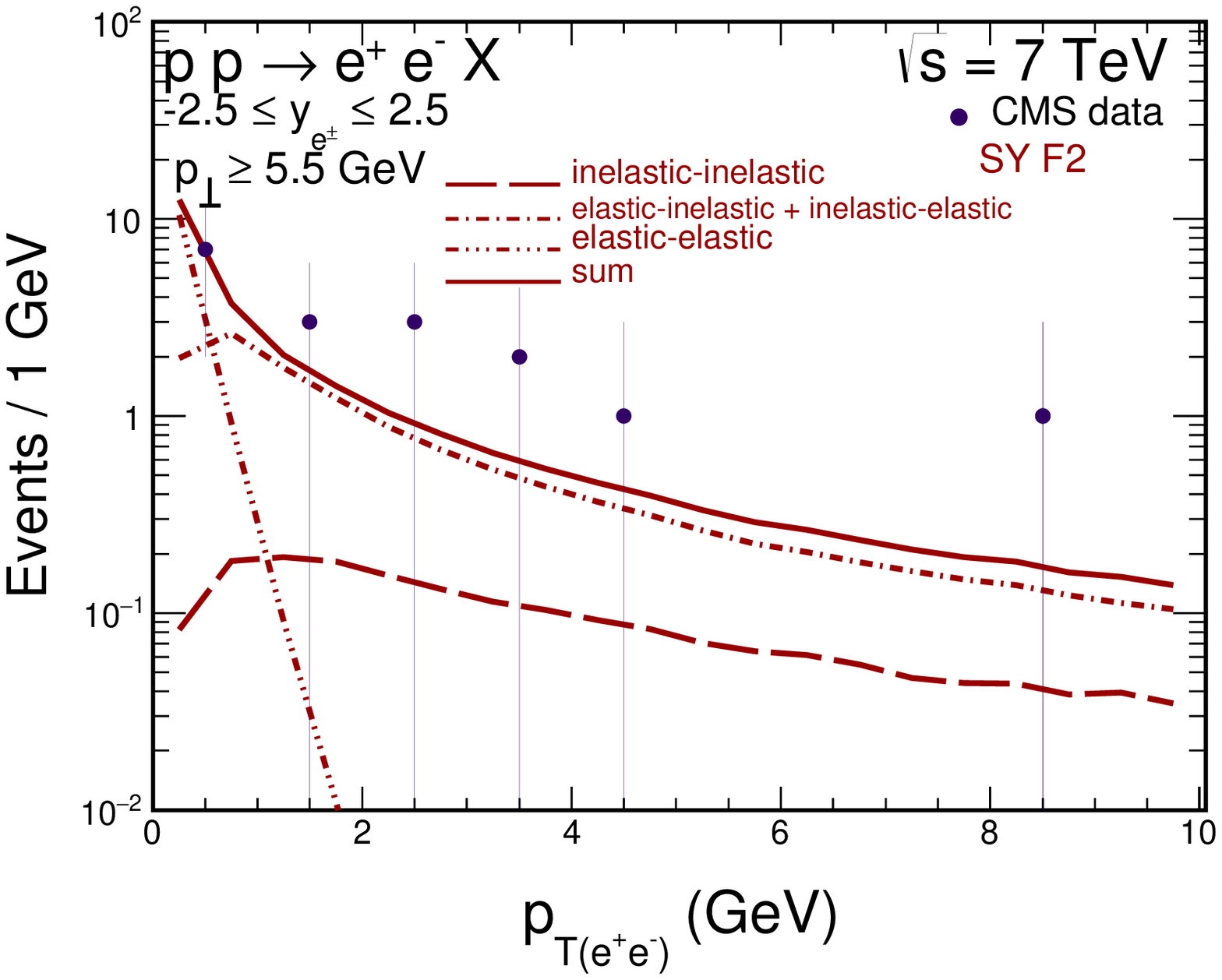}
\includegraphics[width=8cm]{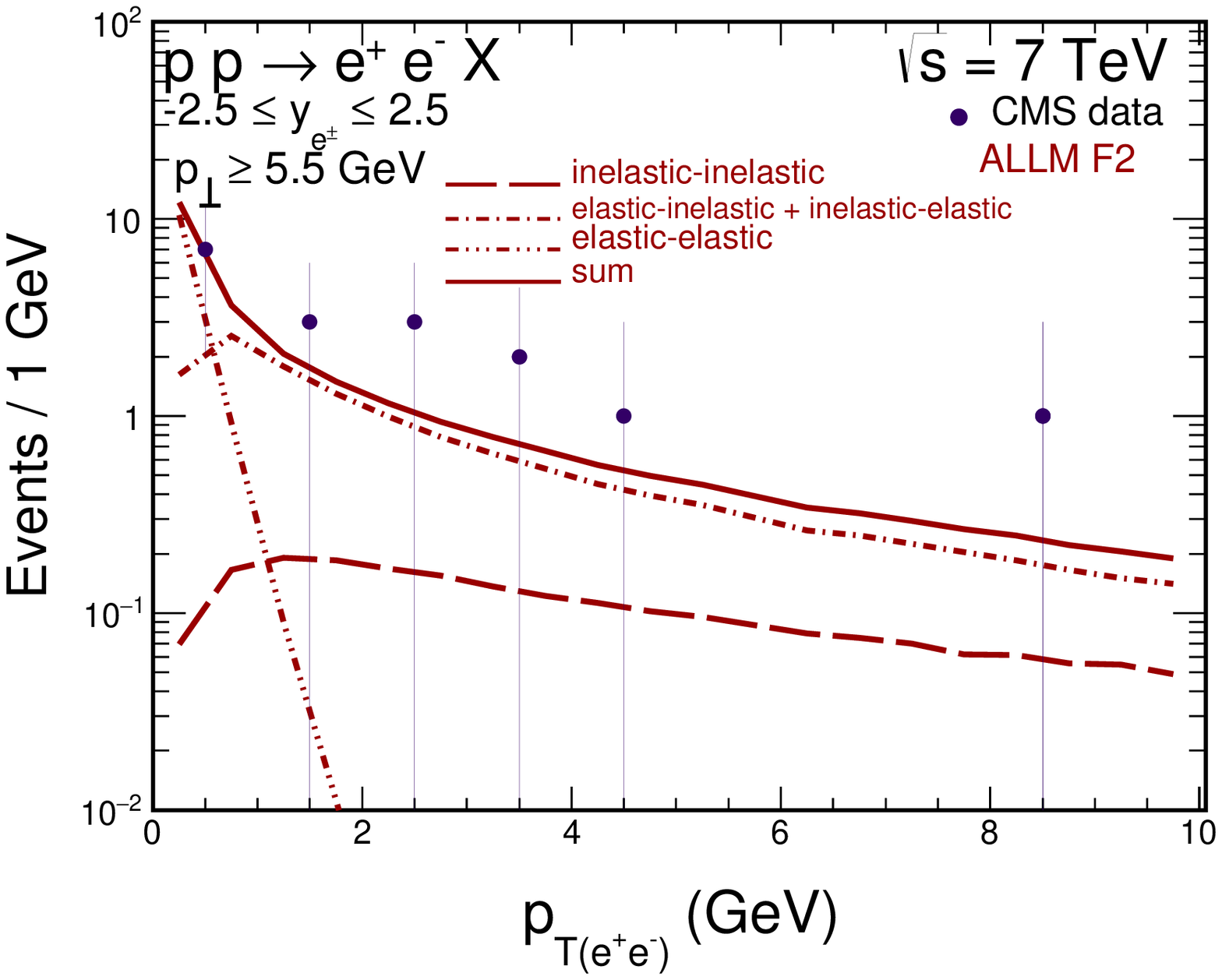}
\end{center}
\caption{
Number of events per pair transverse momentum interval for the CMS
experimental cuts for SY (left) and ALLM (right) structure functions.
The experimental data points are from Ref.\cite{Chatrchyan:2012tv}.
}
\label{fig:CMS_dN_dptsum}
\end{figure}

\begin{figure}
\begin{center}
\includegraphics[width=8cm]{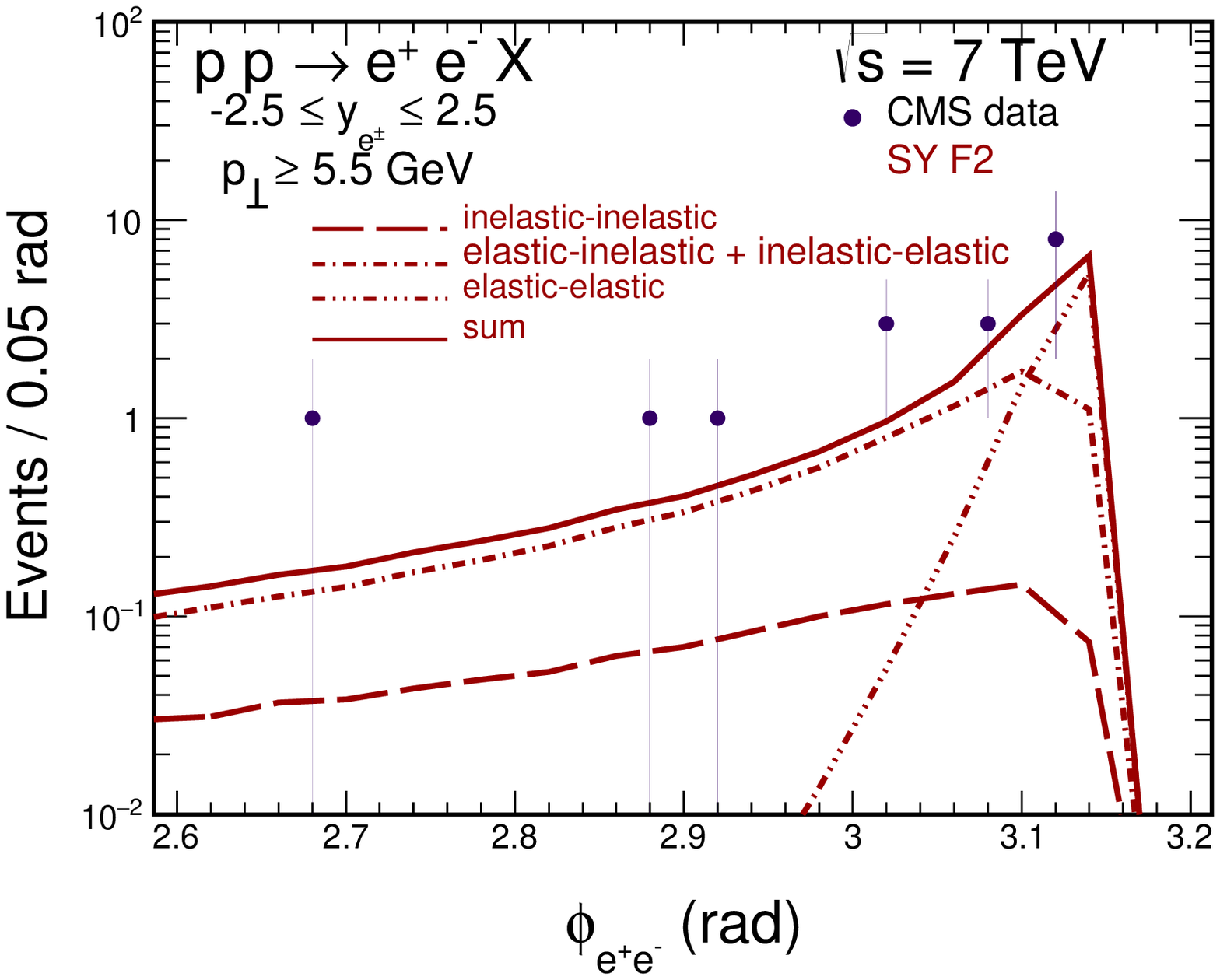}
\includegraphics[width=8cm]{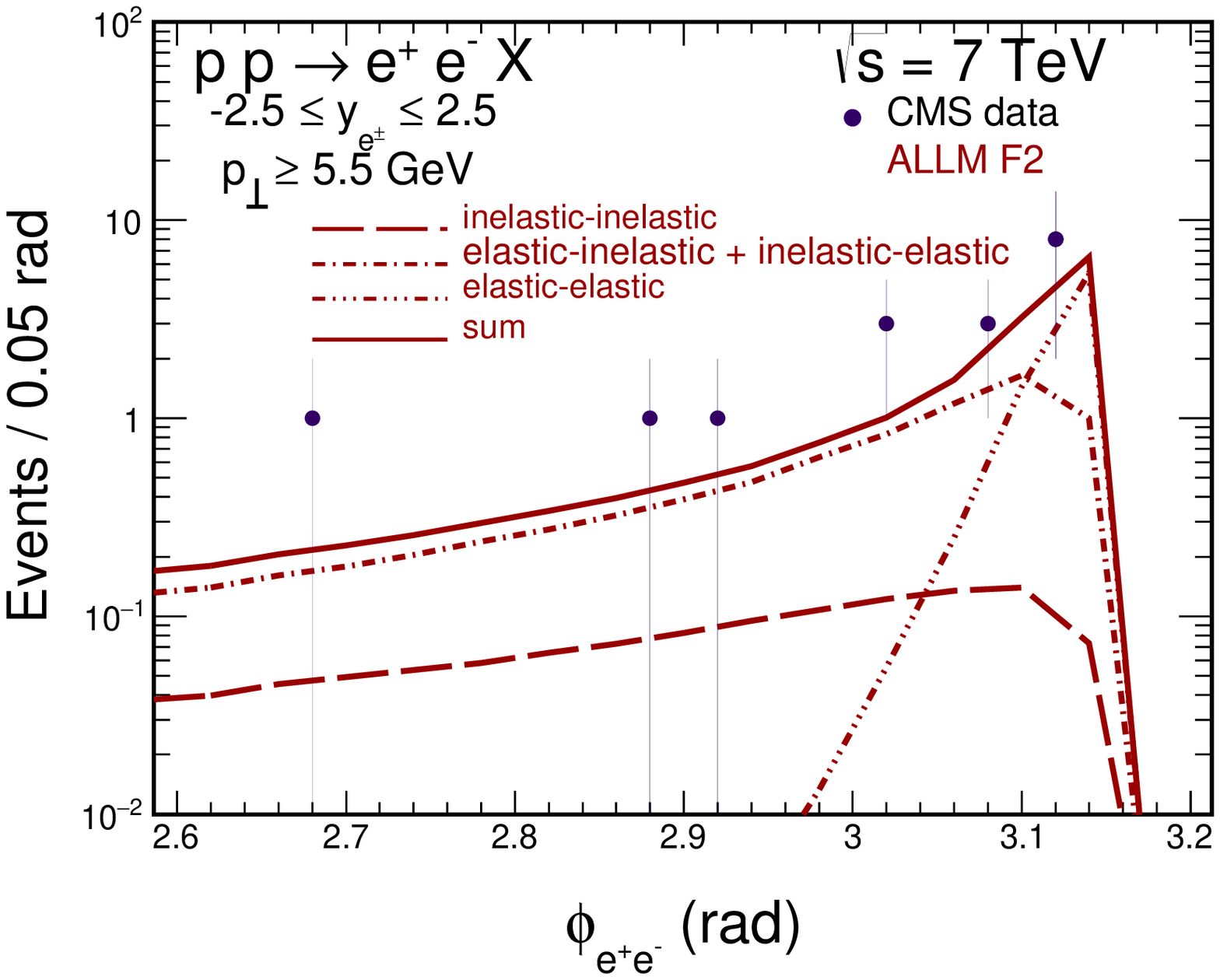}
\end{center}
\caption{
Number of events per pair relative azimuthal angle interval
for the CMS experimental cut for SY (left) and ALLM (right) structure
functions. The experimental data points are from Ref.\cite{Chatrchyan:2012tv}.
}
\label{fig:CMS_dN_dphi}
\end{figure}

In most of the cases considered so far Drell-Yan processes dominate 
\cite{NNS2013,Baranov:2014ewa}.
However, the two-photon processes are interesting by themselves.
Can they be measured experimentally? 
In order to reduce the Drell-Yan contribution and relatively enhance 
the two-photon contribution one can impose an extra condition on lepton
isolation. First trials have been already done by the CMS collaboration
\cite{CMS_gammagamma}. In their analysis an extra lepton isolation
cuts were imposed in order to eliminate the otherwise dominating
Drell-Yan component.
In Figs.
\ref{fig:CMS_dN_dMll},\ref{fig:CMS_dN_dptsum},\ref{fig:CMS_dN_dphi}
we show our results for two different (SY and ALLM) parametrizations 
of the structure functions for distributions in dimuon invariant mass,
in transverse momentum of the pair and in relative azimuthal angle
between $\mu^+ \mu^-$. Only invariant mass distribution
can be obtained in the collinear approach. 
In the collinear approach the second and the third distributions
are just Dirac delta functions in $p_{T,pair}$ and $\phi_{\mu^+ \mu^-}$,
respectively. SY and ALLM parametrizations give almost 
the same contributions to all the distributions considered.
In the first evaluation we have taken into account integrated luminosity
of the experiment ($L$ = 63.2 pb$^{-1}$) as well as experimental
acceptances given in Table 5 in Ref.\cite{Chatrchyan:2012tv}.
Rather good agreement with the low statistics CMS experimental data is
achieved (for both parametrizations of structure functions used in the
figures)
without including any extra corrections due to absorption
effects leading to destroying the rapidity isolation of leptons
and a damping of corresponding cross section
for the photon-photon mechanisms.
This result is interesting by itself.
It may mean that the absorption effects are small or alternatively 
that a contamination of the Drell-Yan contribution is still 
not completely removed.
Both effects should be therefore studied in more detail in a future.
This can be done by full Monte Carlo simulations of both processes 
and clearly goes beyond the scope of the present analysis.

\begin{figure}
\begin{center}
\includegraphics[width=10cm]{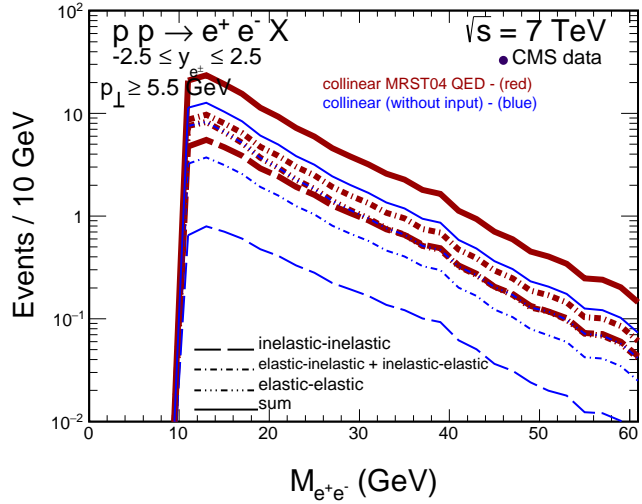}
\end{center}
\caption{
Number of events per invariant mass interval for the CMS experimental
cuts for collinear factorization approach. The results when initial
MRST2004(QED) input is included (thick (red on line) lines) is compared with 
those when it is discarded (thin (blue on line) lines).
The experimental data points are from Ref.\cite{Chatrchyan:2012tv}.
}
\label{fig:CMS_dN_dMll_collinear}
\end{figure}

For completeness and comparison in Fig.\ref{fig:CMS_dN_dMll_collinear} 
we show invariant mass distribution obtained within collinear
factorization approach with $\mu^2 = m_T^2$. We present results for 
the case when initial input
at $Q_0^2$ =  GeV$^2$ (see Eq.(\ref{photon_initial})) is included 
(thick red lines) as well as when it is discarded (thin blue lines)
as discussed in subsection \ref{subsec:DGLAP-photons}.
The results obtained in the letter case are slightly larger than those
obtained within the $k_T$-factorization approach 
(see Fig.\ref{fig:CMS_dN_dMll}), 
especially when the MRST(QED) input is included.

\section{Conclusions}

In the present paper we have discussed in detail production
of dilepton pairs ($e^+ e^-$ or $\mu^+ \mu^-$) in photon-photon
processes in proton-proton scattering at high energies. 
We have compared two different distinct theoretical approaches.

In the first approach photon is treated as a collinear parton
in the proton and included into generalized (QCD,QED) DGLAP equations.
We have discussed and demonstrated that it is not necessary to
keep photon distribution in the evolution equation.
It is sufficient to couple photon to other partons (quarks/antiquarks) 
in the proton,
that undergo usual DGLAP evolution equations.
We have discussed also the issue of initial condition for the
photon distribution at the initial scale. In this context we have 
discussed parametrization/prescription proposed by MRST04(QED) \cite{Martin:2004dh} 
with their initial input as well as when starting evolution
from zero input. The two prescriptions lead to quite different results 
for photon distributions
and in the consequence also for charged lepton observables for finite scales.

In the second approach we take into account the fact that photons 
are off shell and include their transverse momenta and/or virtualities.
We have shown that for typical kinematical conditions of modern
experiments, especially at the LHC, the photon virtualities
are fairly large, which puts doubts on the standard (collinear) 
parton model treatment. The $k_T$-factorization approach
uses unintegrated photon distributions which are expressed
in terms of $F_2$ structure functions \cite{daSilveira:2014jla}. 
Different model parametrizations known from the literature have been 
used in the present study.
The final results depend strongly on the choice of the parametrization.
We have identified regions of the $(Q_i^2, M_i)$ space which give 
significant contribution to the cross section for different experimental 
conditions. For example for the experimental cuts of the recent 
ATLAS experiment \cite{ATLAS_2013_high-mass} mostly perturbative region 
($Q_i^2 >$ 4 GeV and $W_i >$ 3 GeV) contributes. 
Therefore a reliable predictions with accuracy better than 10 \% are possible.
In contrast, for the old ISR \cite{ISR} and more recent PHENIX \cite{PHENIX} 
experiments substantial contributions come from the regions
$W_i <$ 3 GeV and $Q_i^2 <$ 1 GeV. In this case one should 
use explicit parametrizations which fit experimental data in this corner
of the space. The calculation should take into account also
resonance contributions.

In the present paper we have discussed production of dileptons.
Similar analysis may be repeated e.g. for photon-photon induced
production of $W^+ W^-$ pairs. So far only the first approach was
applied there \cite{Luszczak:2014mta}.

\vspace{1cm}


{\bf Acknowledgements}

We would like to thank Juan Rojo for remarks that initiated the present study,
and Laurent Forthomme for help with a Monte Carlo code.
This study was partially supported by the Polish National Science Centre 
grants DEC-2013/09/D/ST2/03724 and DEC-2014/15/B/ST2/02528.


\begin{thebibliography}{299}

\bibitem{Chen:1973mv} 
  M.~S.~Chen, I.~J.~Muzinich, H.~Terazawa and T.~P.~Cheng,
  ``Lepton pair production from two-photon processes,''
  Phys.\ Rev.\ D {\bf 7}, 3485 (1973).
 
\bibitem{Carimalo:1978bu}
  C.~Carimalo, P.~Kessler and J.~Parisi,
  ``$\gamma \gamma$ Background of the {Drell-Yan} Process,''
  Phys.\ Rev.\ D {\bf 18} (1978) 2443
   [Phys.\ Rev.\ D {\bf 19} (1979) 2233].
 
\bibitem{Schrempp:1980zx}
  B.~Schrempp and F.~Schrempp,
  ``Two Photon Exchange in $p$ ($\bar{p}$) $\to \ell^+ \ell^- X$ and a Comparison With {QCD},''
  Nucl.\ Phys.\ B {\bf 182} (1981) 343.
  
\bibitem{daSilveira:2014jla} 
  G.~G.~da Silveira, L.~Forthomme, K.~Piotrzkowski, W.~Sch\"afer and A.~Szczurek,
  ``Central $\mu^{+} \, \mu^{−}$ production via photon-photon fusion in proton-proton collisions with proton dissociation,''
  JHEP {\bf 1502}, 159 (2015)
  [arXiv:1409.1541 [hep-ph]].
 

\bibitem{Budnev:1974de} 
  V.~M.~Budnev, I.~F.~Ginzburg, G.~V.~Meledin and V.~G.~Serbo,
  ``The Two photon particle production mechanism. Physical problems. Applications. Equivalent photon approximation,''
  Phys.\ Rept.\  {\bf 15}, 181 (1975).
 
\bibitem{Vermaseren:1982cz} 
  J.~A.~M.~Vermaseren,
  ``Two Photon Processes at Very High-Energies,''
  Nucl.\ Phys.\ B {\bf 229}, 347 (1983).
  
  
\bibitem{Maciula:2010yw}
R. Maciu{\l}a, A. Szczurek and G. \'Slipek,
``Kinematical correlations of dielectrons from semileptonic decays
of heavy mesons and Drell-Yan processes at BNL RHIC'',
Phys. Rev. {\bf D83} (2011) 054014.

\bibitem{CMS_gammagamma} 
  S.~Chatrchyan {\it et al.} [CMS Collaboration],
  ``Search for exclusive or semi-exclusive photon pair production and observation of exclusive and semi-exclusive electron pair production in $pp$ collisions at $\sqrt{s}=7$ TeV,''
  JHEP {\bf 1211}, 080 (2012)
  [arXiv:1209.1666 [hep-ex]].
%

\bibitem{Ellis:1991qj}
  R.~K.~Ellis, W.~J.~Stirling and B.~R.~Webber,
  ``QCD and collider physics,''
  Camb.\ Monogr.\ Part.\ Phys.\ Nucl.\ Phys.\ Cosmol.\  {\bf 8} (1996) 1.
  
\bibitem{Collins:1984kg}
  J.~C.~Collins, D.~E.~Soper and G.~F.~Sterman,
  ``Transverse Momentum Distribution in Drell-Yan Pair and W and Z Boson Production,''
  Nucl.\ Phys.\ B {\bf 250} (1985) 199.
 

\bibitem{Szczurek:2008ga}
  A.~Szczurek and G.~Slipek,
  ``Parton transverse momenta and Drell-Yan dilepton production,''
  Phys.\ Rev.\ D {\bf 78} (2008) 114007
  [arXiv:0808.1360 [hep-ph]].
 
  
  
\bibitem{NNS2013}
 M.~A.~Nefedov, N.~N.~Nikolaev and V.~A.~Saleev,
  ``Drell-Yan lepton pair production at high energies in the Parton Reggeization Approach,''
  Phys.\ Rev.\ D {\bf 87} (2013) 1,  014022
  [arXiv:1211.5539 [hep-ph]].
  

\bibitem{Baranov:2014ewa}
  S.~P.~Baranov, A.~V.~Lipatov and N.~P.~Zotov,
  ``Drell-Yan lepton pair production at the LHC and transverse momentum dependent quark densities of the proton,''
  Phys.\ Rev.\ D {\bf 89} (2014) 9,  094025
  [arXiv:1402.5496 [hep-ph]].
 

\bibitem{Gluck:2002fi} 
  M.~Gl\"uck, C.~Pisano and E.~Reya,
  ``The Polarized and unpolarized photon content of the nucleon,''
  Phys.\ Lett.\ B {\bf 540}, 75 (2002)
  [hep-ph/0206126].
 

\bibitem{Martin:2004dh} 
  A.~D.~Martin, R.~G.~Roberts, W.~J.~Stirling and R.~S.~Thorne,
  ``Parton distributions incorporating QED contributions,''
  Eur.\ Phys.\ J.\ C {\bf 39}, 155 (2005)
  [hep-ph/0411040].
 

\bibitem{Martin:2014nqa} 
  A.~D.~Martin and M.~G.~Ryskin,
  ``The photon PDF of the proton,''
  Eur.\ Phys.\ J.\ C {\bf 74}, 3040 (2014)
  [arXiv:1406.2118 [hep-ph]].
  

\bibitem{Ball:2013hta} 
  R.~D.~Ball {\it et al.}  [NNPDF Collaboration],
  ``Parton distributions with QED corrections,''
  Nucl.\ Phys.\ B {\bf 877}, 290 (2013)
  [arXiv:1308.0598 [hep-ph]].
  

\bibitem{Schmidt:2014aba} 
  C.~Schmidt, J.~Pumplin, D.~Stump and C.-P.~Yuan,
  ``QED effects and Photon PDF in the CTEQ-TEA Global Analysis,''
  PoS DIS {\bf 2014}, 054 (2014).
 
\bibitem{Luszczak:2014mta}
 M. Luszczak, A. Szczurek and Ch. Royon,
``$W^+ W^-$ pair production in proton-proton collisions:
small missing terms'', 
JHEP {\bf 02} (2015) 098.

\bibitem{Abramowicz:1991xz}
  H.~Abramowicz, E.~M.~Levin, A.~Levy and U.~Maor,
  ``A Parametrization of sigma-T (gamma* p) above the resonance region Q**2 $>=$ 0,''
  Phys.\ Lett.\ B {\bf 269} (1991) 465.
 
\bibitem{Abramowicz:1997ms} 
  H.~Abramowicz and A.~Levy,
  ``The ALLM parameterization of sigma(tot)(gamma* p): An Update,''
  hep-ph/9712415.
 %
\bibitem{Fiore:2002re} 
  R.~Fiore, A.~Flachi, L.~L.~Jenkovszky, A.~I.~Lengyel and V.~K.~Magas,
  ``Explicit model realizing parton hadron duality,''
  Eur.\ Phys.\ J.\ A {\bf 15}, 505 (2002)
  [hep-ph/0206027].
 
\bibitem{Block:2014kza} 
  M.~M.~Block, L.~Durand and P.~Ha,
  ``Connection of the virtual $\gamma^*p$ cross section of $ep$ deep inelastic scattering to real 
$\gamma p$ scattering, and the implications for $\nu N$ and $ep$ 
total cross sections,''
  Phys.\ Rev.\ D {\bf 89}, no. 9, 094027 (2014)
  [arXiv:1404.4530 [hep-ph]].
 
\bibitem{Suri:1971yx} 
  A.~Suri and D.~R.~Yennie,
  ``The Space-time Phenomenology Of Photon Absorption And Inelastic Electron Scattering,''
  Annals Phys.\  {\bf 72}, 243 (1972).
 
\bibitem{Szczurek:1999rd} 
  A.~Szczurek and V.~Uleshchenko,
  ``Nonpartonic components in the nucleon structure functions at small Q**2 in the broad range of x,''
  Eur.\ Phys.\ J.\ C {\bf 12}, 663 (2000)
  [hep-ph/9904288].
 
\bibitem{Pumplin:2002vw}
  J.~Pumplin, D.~R.~Stump, J.~Huston, H.~L.~Lai, P.~M.~Nadolsky and W.~K.~Tung,
  ``New generation of parton distributions with uncertainties from global QCD analysis,''
  JHEP {\bf 0207} (2002) 012
  [hep-ph/0201195].
  
\bibitem{Gehrmann:1999xn}
  T.~Gehrmann, R.~G.~Roberts and M.~R.~Whalley,
  ``A compilation of structure functions in deep inelastic scattering,''
  J.\ Phys.\ G {\bf 25} (1999) A1.

\bibitem{Osipenko:2003ua}
  M.~Osipenko {\it et al.},
  ``The Proton structure function F(2) with CLAS,''
  hep-ex/0309052.
 
\bibitem{Osipenko:2003bu}
  M.~Osipenko {\it et al.} [CLAS Collaboration],
  ``A Kinematically complete measurement of the proton structure function F(2) in the resonance region and evaluation of its moments,''
  Phys.\ Rev.\ D {\bf 67} (2003) 092001
  [hep-ph/0301204].
  
\bibitem{MC_generator}
L.~Forthomme, K.~Piotrzkowski, G.~da~Silveira, W.~Sch\"afer and
A.~Szczurek, to be published.

\bibitem{Root}
{\texttt{https://root.cern.ch/}}

\bibitem{KS2011} 
G.~Kubasiak and A.~Szczurek,
``Inclusive and exclusive diffractive production
of dilepton pairs in proton-proton collisions at high energies'',
Phys. Rev. {\bf D84} (2011) 014005.



\bibitem{PHENIX}
  A.~Adare {\it et al.} [PHENIX Collaboration],
  ``Dilepton mass spectra in p+p collisions at s**(1/2) = 200-GeV and the contribution from open charm,'',
  Phys.\ Lett.\ B {\bf 670} (2009) 313
  [arXiv:0802.0050 [hep-ex]].



\bibitem{LHCb-CONF_2012} 
  [The LHCb collaboration],
  ``Inclusive low mass Drell-Yan production in the forward region at sqrt(s)=7 TeV ,''
  LHCb-CONF-2012-013; Conference report prepared for XX International Workshop on Deep-Inelastic Scattering and Related Subjects, 26-30, March 2012, Bonn, Germany.
  
%
 \bibitem{ATLAS_2014_low-mass} 
  G.~Aad {\it et al.} [ATLAS Collaboration],
  ``Measurement of the low-mass Drell-Yan differential cross section at $\sqrt{s}$ = 7 TeV using the ATLAS detector,''
  JHEP {\bf 1406}, 112 (2014)
  [arXiv:1404.1212 [hep-ex]].  
%
%
\bibitem{ATLAS_2013_high-mass} 
  G.~Aad {\it et al.} [ATLAS Collaboration],
  ``Measurement of the high-mass Drell--Yan differential cross-section in pp collisions at sqrt(s)=7 TeV with the ATLAS detector,''
  Phys.\ Lett.\ B {\bf 725}, 223 (2013)
  [arXiv:1305.4192 [hep-ex]].  
%

\bibitem{CMS:2014jea} 
  V.~Khachatryan {\it et al.} [CMS Collaboration],
  ``Measurements of differential and double-differential Drell-Yan cross sections in proton-proton collisions at 8 TeV,''
  Eur.\ Phys.\ J.\ C {\bf 75}, no. 4, 147 (2015)
  [arXiv:1412.1115 [hep-ex]].
  
\bibitem{ISR}
  C.~Kourkoumelis, L.~K.~Resvanis, T.~A.~Filippas, E.~Fokitis, A.~M.~Cnops, J.~H.~Cobb, R.~Hogue and S.~Iwata {\it et al.},
  ``Study of Massive Electron Pair Production at the CERN Intersecting Storage Rings,''
  Phys.\ Lett.\ B {\bf 91} (1980) 475.  

\bibitem{Chatrchyan:2012tv} 
  S.~Chatrchyan {\it et al.} [CMS Collaboration],
  ``Search for exclusive or semi-exclusive photon pair production and observation of exclusive and semi-exclusive electron pair production in $pp$ collisions at $\sqrt{s}=7$ TeV,''
  JHEP {\bf 1211}, 080 (2012)
  [arXiv:1209.1666 [hep-ex]].
 


\end{thebibliography}
\end{document}